\documentclass[11pt]{book}

\usepackage[dvips]{epsfig}
\usepackage{amssymb}
\usepackage{multirow}
\usepackage{amsbsy}
\usepackage[T1]{fontenc} 
\usepackage[utf8]{inputenc}
\usepackage{amsmath}
\usepackage{xcolor}
\usepackage{amssymb}
\usepackage{mathtools}
\usepackage{amsfonts}
\usepackage{bbold}
\usepackage{enumerate}
\usepackage{tikz}
\usetikzlibrary{angles,quotes}
\usepackage{graphicx}
\usepackage{caption}
\usepackage{subcaption}
\usepackage{float}
\usepackage{ulem}
\usepackage{hyperref}
\usepackage{color}
\usepackage{pstricks}
\catcode`\@=11

 \def\@normalsize{\@setsize\normalsize{13pt}\xipt\@xipt
   \abovedisplayskip 11pt plus3pt minus6pt
   \belowdisplayskip \abovedisplayskip
   \abovedisplayshortskip \z@ plus3pt
   \belowdisplayshortskip 6.6pt plus3.5pt minus3pt}

 \def\small{\@setsize\small{12pt}\xipt\@xipt
   \abovedisplayskip 10pt plus2pt minus5pt
   \belowdisplayskip \abovedisplayskip
   \abovedisplayshortskip \z@ plus3pt
   \belowdisplayshortskip 6pt plus3pt minus3pt
   \def\@listi{\topsep 6pt plus 2pt minus 2pt
     \parsep 3pt plus 2pt minus 1pt
     \itemsep \parsep}}

 \def\footnotesize{\@setsize\footnotesize{10pt}\ixpt\@ixpt
   \abovedisplayskip 8pt plus 2pt minus 4pt
   \belowdisplayskip \abovedisplayskip
   \abovedisplayshortskip \z@ plus 1pt
   \belowdisplayshortskip 4pt plus 2pt minus 2pt
   \def\@listi{\topsep 4pt plus 2pt minus 2pt
      \parsep 2pt plus 1pt minus 1pt
      \itemsep \parsep}}

 \def\scriptsize{\@setsize\scriptsize{9.5pt}\viiipt\@viiipt}
 \def\tiny{\@setsize\tiny{7pt}\vipt\@vipt}
 \def\large{\@setsize\large{14pt}\xiipt\@xiipt}
 \def\Large{\@setsize\Large{18pt}\xivpt\@xivpt}
 \def\LARGE{\@setsize\LARGE{22pt}\xviipt\@xviipt}
 \def\huge{\@setsize\huge{25pt}\xxpt\@xxpt}
 \def\Huge{\@setsize\Huge{30pt}\xxvpt\@xxvpt}
\def\section{\@startsection {section}{1}{\z@}%
{-1.5\baselineskip plus-1pt minus-3pt}{1\baselineskip plus1pt minus2pt}%
{\centering\normalsize\bf}}
\def\subsection{\@startsection{subsection}{2}{\z@}%
{-1\baselineskip plus-1pt minus-2pt}{1\baselineskip plus1pt minus2pt}%
{\normalsize\sc\noindent}}
\def\subsubsection{\@startsection{subsubsection}{3}{\z@}%
{-1\baselineskip plus-1pt minus-2pt}{1sp}{\normalsize\it\noindent}}
\def\paragraph{\@startsection{paragraph}{4}{\z@}%
{1\baselineskip plus1pt minus2pt}{-1em}{\normalsize\it\noindent}}
\let\subparagraph=\paragraph

\def\tableofcontents{\@restonecolfalse\if@twocolumn\@restonecoltrue
\onecolumn\fi\OSIDcont\@starttoc{con}\if@restonecol\twocolumn\fi}

\def\l@section{\@dottedtocline{1}{0em}{.66em}}

\def\thebibliography#1{\section*{{Bibliography}\@mkboth
 {BIBLIOGRAPHY}{BIBLIOGRAPHY}}\footnotesize\rm\list
 {[\arabic{enumi}]}{\settowidth\labelwidth{[#1]}\leftmargin\labelwidth
 \advance\leftmargin\labelsep\usecounter{enumi}}
 \def\newblock{\hskip .11em plus .33em minus -.07em}
 \sloppy\clubpenalty4000\widowpenalty4000
 \sfcode`\.=1000\relax}

\def\ps@myheadings{\let\@mkboth\@gobbletwo
\def\@oddhead{\hfil{\footnotesize\rm\rightmark}\hfil}
\def\@evenhead{\hfil{\footnotesize\rm\leftmark}\hfil}
\def\@oddfoot{\hfil{\footnotesize\sf\artid-\thepage}\hfil}
\def\@evenfoot{\hfil{\footnotesize\sf\artid-\thepage}\hfil}
\def\sectionmark##1{}\def\subsectionmark##1{}}

\def\@copyrighthead{\parbox{127mm}{\footnotesize\rm\ \\[6pt]
Open Systems~\& Information Dynamics\\
Vol.~\Vol, No.~\Number~(\Year)~\artid~(\EndpagE~pages)\\
DOI:\DOInumber\\
\copyright~World Scientific Publishing Company\\
\epsfxsize=4cm
\vskip-\lastskip
\vskip-\baselineskip
\vspace*{-38.5pt}
\noindent\hfill\epsfbox{wlogo.eps}}}

\def\artid{0000001}
\def\Year{2008}        %
\def\Vol{15}           % <-------( ustawienia robocze )
\newcounter{paPer}     %
\def\EndpagE{\expandafter\pageref{\the\value{paPer}OpSy}}

\def\ps@osiD{\let\@mkboth\@gobbletwo
\def\@oddhead{\@copyrighthead}
  \def\@oddfoot{\hfil{\footnotesize\sf\artid-\thepage}\hfil}
  \def\@evenhead{}\let\@evenfoot\@oddfoot}

\def\cite{\@ifnextchar [{\@tempswatrue\@Rcitex}{\@tempswafalse\@Rcitex[]}}

\def\@Rcitex[#1]#2{\if@filesw\immediate\write\@auxout{\string\citation{#2}}\fi
  \def\@citea{}\@cite{\@for\@citeb:=#2\do
    {\@citea\def\@citea{,\penalty\@m\,}\@ifundefined
       {b@\the\value{paPer}R\@citeb}{{\bf ?}\@warning
       {Citation `\@citeb' on page \thepage \space undefined}}%
\hbox{\csname b@\the\value{paPer}R\@citeb\endcsname}}}{#1}}

\long\def\@caption#1[#2]#3{\par\addcontentsline{\csname
  ext@#1\endcsname}{#1}{\protect\numberline{\csname
  the#1\endcsname}{\ignorespaces #2}}\begingroup
    \@parboxrestore
    \small                                        %    \normalsize
    \@makecaption{\csname fnum@#1\endcsname}{\ignorespaces #3}\par
  \endgroup}

\newtoks\@stequation

\def\subequations{\refstepcounter{equation}%
\edef\@savedequation{\the\c@equation}%
\@stequation=\expandafter{\theequation}%   %only want \theequation
\edef\@savedtheequation{\the\@stequation}% %expanded once
\edef\oldtheequation{\theequation}%
\setcounter{equation}{0}%
\def\theequation{\oldtheequation\alph{equation}}}%

\def\endsubequations{%
\setcounter{equation}{\@savedequation}%
\@stequation=\expandafter{\@savedtheequation}%
\edef\theequation{\the\@stequation}\global\@ignoretrue}

\catcode`\@=12
\let\Rlabel=\label
\let\Rbibitem=\bibitem
\let\Rref=\ref
\let\Rpageref=\pageref
\def\label#1{\expandafter\Rlabel{\the\value{paPer}R#1}}
\def\bibitem#1{\expandafter\Rbibitem{\the\value{paPer}R#1}}
\def\ref#1{\expandafter\Rref{\the\value{paPer}R#1}}
\def\pageref#1{\expandafter\Rpageref{\the\value{paPer}R#1}}

\def\thesection{\arabic{section}.}

\def\YYMm{\rule{0ex}{4em}}
\newtoks\TITsi
\newtoks\TITsii

\def\title#1{\def\TITs{\LARGE{\raggedright\noindent\YYMm #1%
\vskip8pt\par}}}
\def\staR{$^{\,*}$}

\def\author#1{\autMM{#1}\def\LHD{#1}}
\def\and{{\rm\lowercase{and}}}

\def\autMM#1{\TITsii={\vskip10pt\par\normalsize\rm\noindent #1\par}%
\TITsi=\expandafter{\TITs}\edef\TITs{\the\TITsi\the\TITsii}}

\def\address#1{\TITsii={\vskip6pt\par\footnotesize\sl\noindent #1\par}%
\TITsi=\expandafter{\TITs}%
\edef\TITs{\the\TITsi\the\TITsii}}

\def\received#1{\TITsii={\vskip10pt\par\small\rm\noindent(Received: #1)\par}%
\TITsi=\expandafter{\TITs}\edef\TITs{\the\TITsi\the\TITsii}}

\def\headtitle#1{\def\RHD{#1}}
\def\headauthor#1{\def\LHD{#1}}
\def\listas#1#2{\addcontentsline{con}{section}{{\sc #1: }{\rm #2}}}

\def\abst{{\bf Abstract.}}
\def\abstract#1{\TITs
       \vskip15pt\par\noindent
       {\footnotesize{\abst~} #1\vskip3pt\par}
       \markright{\RHD}
       \markboth{\LHD}{\RHD}}

\def\OSIDcont{\cleardoublepage\thispagestyle{empty}
       \markright{}\markboth{}{}
       \normalsize\rm
       \hspace*{\fill}{\large\rm
         Contents of the Volume \Volume, Number \Number}\hspace*{\fill}
       \par\vspace{1.5em}
       \par\noindent}

\def\endpaper{\expandafter\label{\the\value{paPer}OpSy}}

\def\1{{\mathchoice{\rm 1\mskip-4mu l}{\rm 1\mskip-4mu l}%
{\rm 1\mskip-4.5mu l}{\rm 1\mskip-5mu l}}}
\def\varkappa{\mbox{\bBB\char 123}}
\def\longhookrightarrow{\lhook\joinrel\relbar\joinrel\rightarrow}

\def\longhookUp{\lower6pt\hbox{\rotatebox{90}{$\longhookrightarrow$}}}

\def\theequation{\thesection\arabic{equation}}

\def\Myskip{\setlength{\baselineskip}{13pt}}
\def\text#1{\quad\mbox{\rm  #1 }\quad}

\def\DOInumber{}
\def\SMP{\def\thefootnote{*}%
\def\thefootnote{\arabic{footnote}}\setcounter{footnote}{0}}

\input xy
\xyoption{all}

\InputIfFileExists{psfig.sty}{\typeout{^^Jpsfig.sty inputed...ok}}{\typeout{^^JWarning: psfig.sty could not be found.^^J}}
\def\artid{0000001}
\def\Volume{15}
\def\Number{1}
\def\Year{2008}
\def\DOInumber{}

\begin{document}

\newcommand{\Mn}{M_n(\mathbb{C})}
\newcommand{\Mk}{M_k(\mathbb{C})}
\newcommand{\id}{\mbox{id}}
\newcommand{\ot}{{\,\otimes\,}}
\newcommand{{\Cd}}{{\mathbb{C}^d}}
\newcommand{\sbsigma}{{\mbox{\scriptsize \boldmath $\sigma$}}}
\newcommand{\sbalpha}{{\mbox{\scriptsize \boldmath $\alpha$}}}
\newcommand{\sbbeta}{{\mbox{\scriptsize \boldmath $\beta$}}}
\newcommand{\bsigma}{{\mbox{\boldmath $\sigma$}}}
\newcommand{\balpha}{{\mbox{\boldmath $\alpha$}}}
\newcommand{\bbeta}{{\mbox{\boldmath $\beta$}}}
\newcommand{\bmu}{{\mbox{\boldmath $\mu$}}}
\newcommand{\bnu}{{\mbox{\boldmath $\nu$}}}
\newcommand{\ba}{{\mbox{\boldmath $a$}}}
\newcommand{\bb}{{\mbox{\boldmath $b$}}}
\newcommand{\sba}{{\mbox{\scriptsize \boldmath $a$}}}
\newcommand{\MD}{\mathfrak{D}}
\newcommand{\sbb}{{\mbox{\scriptsize \boldmath $b$}}}
\newcommand{\sbmu}{{\mbox{\scriptsize \boldmath $\mu$}}}
\newcommand{\sbnu}{{\mbox{\scriptsize \boldmath $\nu$}}}
\def\oper{{\mathchoice{\rm 1\mskip-4mu l}{\rm 1\mskip-4mu l}%
{\rm 1\mskip-4.5mu l}{\rm 1\mskip-5mu l}}}
\def\<{\langle}
\def\>{\rangle}
\def\theequation{\thesection\arabic{equation}}

\title{Effective dynamics of qubit networks via phase-covariant quantum ensembles\staR}
\author{Sean Prudhoe, Unnati Akhouri, Tommy Chin, and Sarah Shandera}
\address{Institute of Gravitation and the Cosmos, The Pennsylvania State University,\\
Whitmore 321, University Park, PA, USA}
\address{Department of Physics, The Pennsylvania State University,\\
Davey 104, University Park, PA, USA}
\headauthor{}
\headtitle{Effective dynamics of qubit networks}
\received{May 1, 2024}
\listas{Sean Prudhoe, Unnati Akhouri, Tommy Chin, and Sarah Shandera}{Effective dynamics of qubit networks via phase-covariant quantum ensembles}

\abstract{We derive a new constructive procedure to rapidly generate ensembles of phase-covariant dynamical maps that may be associated to the individual spins of a closed quantum system. We do this by first computing the single-spin dynamical maps in small XXZ networks and chains, specialized to the class of initial states that guarantees phase-covariant dynamics for each spin. Since the dynamics in any small, closed system contains oscillatory features associated to the system size, we define an averaging procedure to extract time-homogeneous dynamics. We use the the average map and the set of deviations from the average map in the exactly derived ensembles to motivate the form of distributional functions for map parameters. The distributions then straightforwardly generate arbitrary-sized ensembles of channels, constrained by a few global properties. This procedure can also generate ensembles where individual maps are not phase-covariant although the average map is, corresponding to realizations of disordered, or noisy, Hamiltonians. The construction procedure suggests new ways to realize random families of open-system dynamics, subject to constraints that require the ensemble to approximate a partition of a closed system.}

%{We study ensembles of phase-covariant channels. We show that such ensembles arise naturally from familiar spin-chain models (e.g., XXZ) with a special class of initial states, and that the disorder-averaged map of disordered spin chains is phase-covariant under a weak symmetry constraint on the distribution. We use those examples to motivate a broader class of phase-covariant ensembles, which include both unital and non-unital channels. We demonstrate the physical properties captured by the late-time limit of the average map over the ensemble.}

\Myskip

%\pacs{03.65.Ud, 03.67.-a}

\SMP

\section{Introduction}
\setcounter{equation}{0}
Most of the quantum systems used for measurement and computation are open, rather than isolated, systems. We are frequently interested in understanding the dynamics of an ensemble of such systems, such as the set of evolutions of some quantum state under many trials of a noisy circuit \cite{Deshpande:2022,Fefferman:2023}. An ensemble of noisy circuits is an approximate description of some more complex, closed system, and more generally ensembles of open systems provide an alternative description of closed systems. In that case, the ensemble is generated by a partition of the closed system into a set of subsystems, with the evolution of each subsystem described by an open-systems equation. The set of sub-system evolutions is constrained by conservation laws and symmetries of the closed, unitary evolution. One might in practice be primarily interested only in the evolution of some special subsystems, for example the computational qubits or the probe systems for a measurement task. For other questions, such as studying thermalization, it may be useful to define a typical or average open evolution together with the deviations from the average.  

In this work, we derive and characterize ensembles of open-system evolutions where either all members of the ensemble are constrained by a symmetry (here, phase-covariance \cite{Holevo:1993,Lankinen:2016}, defined in detail below), or where the ensemble-averaged map carries the symmetries even though the individual maps may not. We explicitly calculate the dynamical maps associated to individual spins for several small systems. While the individual maps are characterized by parameters with oscillitory behavior in time, we provide a prescription to define a time-independent average map. We then propose distributions for the parameters in a dynamical map, tying the distribution parameters to properties of the closed-system ensembles. This is a qualitatively new approach to ensembles of open systems.

Prior work on random ensembles of open-system dynamics falls into a few classes. Some early work includes studies of dynamics produced from Haar random unitaries acting on a larger, system + environment, Hilbert space, and dynamics of systems subjected to periodic measurement \cite{Bruzda:2010}. A number of recent studies \cite{Denisov:2019,Can:2019b,Can:2019,Lieu:2020,Tarnowski_2023} consider random Lindbladian master equations, which restricts the open-system evolution to be time-homogeneous and Markovian. Related approaches include studies of random non-Hermitian Hamiltonians \cite{Li:2021}, which arise from an approximate truncation of the Lindblad master equation, and random Kraus maps \cite{Sa:2020}, which are also restricted to Markovian dynamics. Studies of noise in quantum circuits consider ensembles of dissipative maps, but traditionally focus on specific unital channels (e.g., amplitude damping or depolarization channels). 

This prior work focuses on random dynamics characterized only by loose restrictions (i.e., Lindblad form, or amplitude damping), generally disconnected from the internal dynamics of the open system itself. In contrast, we are interested in characterizing open-system dynamics for the subsystems that together make up a closed system with some particular properties. As a first step, we consider a simple class of symmetry-restricted dynamics, phase-covariant maps. These depend on just a few parameters, but provide a well-characterized way to consider open-system evolution that may be non-Markovian, non-unital, and time-inhomogeneous \cite{Filippov:2019,https://doi.org/10.48550/arxiv.2106.05295,Siudzinska:2023a} which makes them appealing for phenomenological studies of thermalization, quantum homogenization, dephasing processes, quantum metrology, and quantum optics \cite{Smirne2016,Haase:2018,Haase2019,Filippov:2019}. The non-unitality is especially of interest in the context of quantum computation and information \cite{Fefferman:2023,Siudzinska:2023b,Siudzinska:2023c}. 

The most general phase-covariant qubit channel can be parameterized as 
\begin{equation}
\label{eqn::PhaseCovMap}
\Lambda_{\rm PC}=\begin{bmatrix} 1&0&0&0 \\ 0&\lambda_{1}\cos\theta&-\lambda_{1}\sin\theta&0 \\ 0&\lambda_{1}\sin\theta&\lambda_{1}\cos\theta &0\\ \tau_{3}&0&0&\lambda_{3}\end{bmatrix} \,,
\end{equation}
where $\lambda_1$, $\lambda_3$, $\theta$, and $\tau_3$ are all real numbers and complete positivity imposes some restrictions on the parameters, discussed below. This matrix acts on the vector characterizing the density matrix in the Pauli basis. That is, using the Pauli matrices $X,Y,Z$, writing $\rho=\frac{1}{2}(\mathbb{1}+a_xX+a_yY+a_zZ)$, and defining $\vec{v}_{\rho_i}=(1,a_x,a_y,a_z)^{T}$, the map generates a new state defined by the components of the vector $\vec{v}_{\rho_f}=\Lambda_{\rm PC}\vec{v}_{\rho_i}$. (Our convention for the Pauli matrices is given in Eq.(\ref{eq:Paulis})). 

The action of this class of channels includes translations and $xy$-symmetric deformations of the Bloch sphere that commute with rotations about the $z$-axis. Phase covariant maps provide agnostic noise models, since they describe a combination of pure dephasing with energy absorption and emission. An additional useful property of phase-covariant channels is that they need not have the maximally mixed state as a fixed point. That is, they need not be unital. The state that is invariant under the action of the phase-covariant map, Eq.(\ref{eqn::PhaseCovMap}), is $\rho_* =  \frac{1}{2}\left(\mathbb{1}+ \left(\frac{\tau_3}{1-\lambda_{3}}\right) Z\right)$.

 This state is different from the maximally mixed state (the Gibbs state at infinite temperature) whenever $\tau_3\neq 0$, but it is still diagonal and can be understood as a classical ensemble or Gibbs state, $\rho\propto e^{-\beta_* H}$ with $\beta_* = \log \Big[ \frac{2}{1-a_*}\Big]>0 $, where $a_* = \frac{\tau_3}{1-\lambda_3}$. This simple steady state is useful for studying thermalization beyond the Markovian regime \cite{Lankinen:2016}. 

Evolution under a phase-covariant dynamical map (a smooth, time parameterized sequence of phase-covariant channels) can be generated for a system with free Hamiltonian $H_{\rm S}$ by coupling it via an interaction term $H_{\rm SE}$ to an environment with free Hamiltonian $H_{\rm E}$ in initial state $\rho_{\rm S}\otimes\rho_{\rm E}$, with the restrictions 
\begin{equation}
\label{eq:PCcond}
   [H_{\rm S}+H_{\rm E},H_{\rm SE}]=[H_{\rm E},\rho_{\rm E}]=0\,.
\end{equation}
These relations imply that (i) no energy builds up within interactions and (ii) the initial environment state is a Gibbs state with respect to $H_{\rm E}$, and plays the role of a classical reservoir in the case that the reduced dynamics is Markovian. But, these maps are generically non-Markovian, allowing much richer dynamics \cite{Chruscinski:2022}.

By choosing Hamiltonians for $N$ coupled spins with initial states that obey the conditions above, we can construct ensembles of open-system evolution where each member of the ensemble is a phase-covariant map for a single spin degree of freedom. Without translation symmetry for example, the full set of maps is time-consuming to derive for large $N$, as each 1-qubit dynamical map needs to be computed explicitly. Leveraging symmetries, such as translation symmetry, we are able to compute all 1-qubit dynamical maps through a single computation, thus allowing computation of the network average. 

 We find that certain averaged dynamical maps provide a notion of stationary channels that depends only on the network initial state and the symmetries of the interaction terms in the Hamiltonian. To construct these channels from example systems, we first average over the set of individual dynamical maps at fixed time $t$ associated with the evolution of the $i$th subsystem, $\Lambda_i(t')$,  and then take the limit of a long time-average:
\begin{align}
\label{eq:LamdaTensembleAve}
\overline{\Lambda}^{\infty}_{N}\equiv\lim_{t\rightarrow\infty} \frac{1}{t}\int_{0}^{t}\left[\frac{1}{N}\sum_{i=1}^{N
    }\Lambda_{i}(t')\right]dt'\,.
\end{align}
For any finite $N$, the order in which these limits are taken is immaterial, the same result is obtained in either case. However, this may not hold for the case with infinitely many spins. In that case one must prove that the summation over the dynamical maps converges absolutely with respect to some appropriately defined norm (Fubini's Theorem). While such a result seems plausible, a proof lies beyond the scope of this paper. Regardless, we find a physical motivation to take the averages in the order given in Eq.(\ref{eq:LamdaTensembleAve}), discussed in further in Section 4. 

In useful cases, the limit of the long-time and ensemble averaged map will be a time-independent channel. Furthermore, for disordered Hamiltonians, $\overline{\Lambda}_{N}(t)$ (and the map averaged over the $N$ sub-systems at any time) may be phase-covariant even when individual maps are not. This is most directly seen by working with the distribution of map parameters, inherited from the Hamiltonian disorder, over the $N$ sub-systems. Then $\overline{\Lambda}^{\infty}_{N}$ may again be a time-independent, phase-covariant channel. We will demonstrate these points with specific examples below.

In the rest of the paper, we establish a procedure for generating ensembles of phase-covariant maps with a long-time average that is time-independent and with fluctuations about the average that mimic those found in finite-size spin networks. We begin in Section 2 by constructing ensembles of maps for spin networks where every division into sub-system and environment satisfies Eq.(\ref{eq:PCcond}). That is, every map is phase-covariant and differences in the map come from the inhomogeneity of the initial state. We derive the $\overline{\Lambda}^{\infty}_{N}$ for those ensembles. In Section 3 we provide a perturbative calculation justifying the use of $\overline{\Lambda}^{\infty}_{N}$ as defined in Eq.(\ref{eq:LamdaTensembleAve}) as a large-$N$ limit for at least some kinds of systems. Next, in Section 4, we consider a case where different members of the ensemble may also have different dynamics, for example as generated by different values for couplings within the full, closed Hamiltonian. There we demonstrate how moving from a distribution over Hamiltonian parameters to a distribution over map parameters facilitates the construction of additional scenarios whose average dynamics are phase covariant.  We put these pieces together in Section 5, and provide a prescription for generating ensembles of open-system evolution that we conjecture to be useful for studying the ensemble of subsystems of large-$N$, closed systems. We conclude with discussion in Section 6.

%%%%%%%%%%%%%%%%%%%%%%%%%%%%%%%%%%%%%%%%%%%%%%%%%
 %%%%%%%%%%%%%%%%%%%%%%%%%%%%%%%%%%%%%%%%%%%%%%%%%
\section{Phase-covariant ensembles from XXZ networks} \label{sec:physcons}
\setcounter{equation}{0}
%%%%%%%%%%%%%%%%%%%%%%%%%%%%%%%%%%%%%%%%%%%%%%%%%
%%%%%%%%%%%%%%%%%%%%%%%%%%%%%%%%%%%%%%%%%%%%%%%%%
In this section we consider several examples of the dynamical maps appearing in spin-networks, which we use as benchmark examples of the phase-covariant ensembles. We are interested in extracting features that can be usefully generalized to a construction that can be carried out using only the parameters of the maps in Eq.(\ref{eqn::PhaseCovMap}), without reference to an explicit Hamiltonian, although we will see how properties of the Hamiltonian are recognizable in the maps. The phase-covariant restrictions on the examples we consider lead to a simple late-time averaged behavior, described by a single phase-covariant channel (since these channels are closed under convex linear combinations \cite{Siudzinska:2022}). Besides characterizing this channel, we use these models to extract a typical scaling with $N$ of how maps vary around the average, which can be used to inform more general probability distributions over map parameters.  

To begin, the dynamical map associated to the $k^{\rm th}$ qubit in the network is defined as
\begin{equation}
\begin{split}
\label{eq:DMviatrace}
      \Lambda_{k}(t)[\rho_{k}(0)]&={\rm tr}_{j\neq k}\left[e^{-iHt}\rho_{k}(0)\otimes\rho_{\rm E_{k}}(0)e^{iHt}\right]
\end{split}
 \end{equation}
 where $H$ is the full Hamiltonian describing the network dynamics. $\Lambda_{k}$(t) depends on time, the Hamiltonian parameters, and the initial state of the remaining qubits. To obtain phase-covariant dynamical maps for a given qubit network, we must ensure the thermodynamic constraints from Eq.(\ref{eq:PCcond}) are satisfied. Therefore, we choose initial global states that are totally uncorrelated and diagonal in the computational basis (as appropriate for the Hamiltonians we consider below)
\begin{equation}
\label{eq:IC}
\rho(0)=\frac{1}{2^{N}}\bigotimes_{k=1}^{N}\big(\mathbb{1}_{k}+z_{k}Z_{k}\big) \,,
\end{equation}
Where $N$ is the number of qubits (or spins) within the network and $Z_{k}$ is the third Pauli matrix associated to the $k^{th}$ qubit's Hilbert space. The Bloch vectors for each qubit are initially parallel to the $z$-axis with length $z_{k}$. The initial state must be totally uncorrelated so that for all possible choices of focal qubit the reduced dynamics is completely positive. Then, all maps are of the form Eq.(\ref{eqn::PhaseCovMap}) and satisfy the following inequalities \cite{Ruskai2001}
 \begin{equation}
 \label{eq:CP2}
\begin{split}
|\lambda_{3,k}|+|\tau_{3,k}|&\leq1\\
4\lambda^{2}_{1,k}+\tau^{2}_{3,k} &\leq(1+\lambda_{3,k})^{2}\,.
     \end{split}
 \end{equation}

%%%%%%%%%%%%%%%%%%%%%%%%%%%%%%%%%%%%%%%%%%
%%%%%%%%%%%%%%%%%%%%%%%%%%%%%%%%%%%%%%%%%%
\subsection{XXZ-networks}
%%%%%%%%%%%%%%%%%%%%%%%%%%%%%%%%%%%%%%%%%%
%%%%%%%%%%%%%%%%%%%%%%%%%%%%%%%%%%%%%%%%%%
In this section, we consider homogeneous XXZ-networks with two different topologies. This terminology refers to spin networks with two-qubit interactions, where the coefficients of the $X_{i}X_{j}$ and $Y_{i}Y_{j}$ terms are the same, while the coefficient of $Z_{i}Z_{j}$ is different. The first class we consider has nearest-neighbor interactions and periodic boundary conditions (a ring), with Hamiltonian 
\begin{equation}
    H^{\rm ring}_{\rm XXZ}=h\sum_{i=1}^{N}Z_{i}+\frac{J_{\perp}}{2}\sum_{i=1}^{N}(X_{i}X_{i+1}+Y_{i}Y_{i+1})+\frac{J_{\parallel}}{2}\sum_{i=1}^{N}Z_{i}Z_{i+1}\,.
\end{equation}
Here $X,Y,Z$ are the usual Pauli matrices and the subscript index indicates which spin the operator acts on (tensor products with the identity on all others is assumed). The indices take integer values mod $N$. The coupling constants $J_{\parallel}$ and $J_{\perp}$ determine the class of the model, and the parameter $h$ is the magnetization. For $J_{\perp}=0$, $H_{\rm XXZ}$ reduces to the (anti)-ferromagnetic Ising model with $J_{\parallel}<0$ ($J_{\parallel}>0$). When $J_{\perp}=J_{\parallel}$ the model reduces to the XXX model. These models belong to  different universality classes as they flow to distinct fixed points under renormalization. 

Second, we consider completely connected networks, with all possible pairwise interactions:
\begin{equation}
    H^{cc}_{\rm XXZ}=h\sum_{i=1}^{N}Z_{i}+\frac{J_{\perp}}{2}\sum_{i,j}(X_{i}X_{j}+Y_{i}Y_{j})+\frac{J_{\parallel}}{2}\sum_{i,j}Z_{i}Z_{j} \,.
\end{equation}
We choose these two cases because they are highly symmetric, without any boundary spins and with all spins having the same number and type of interactions.

Technically, the computation of the dynamical maps via Eq.(\ref{eq:DMviatrace}) involves exponentiating the Hamiltonian and tracing over most of the system. This is easiest to do if the Hamiltonian is first diagonalized. Both topologies have a number of symmetries that can be exploited to perform this diagonalization. A detailed description can be found in the Appendix, but we summarize the key points here as they provide a helpful organization to understand some of the structure in the maps.  

First, $H_{\rm XXZ}$ for either connectivity has a U(1) symmetry generated by the charge operator $Q_{N}=\sum_{i}Z_{i}$. Therefore $H_{\rm XXZ}$ is block-diagonal in an excitation basis,
\begin{equation}
\mathcal{P}_{N}H_{\rm XXZ}\mathcal{P}_{N}^{\dag}=H^{0}_{\rm XXZ}\oplus H^{1}_{\rm XXZ}\oplus...\oplus H^{N}_{\rm XXZ}\,,
\end{equation}
with dim $H^{q}_{\rm XXZ}= {N\choose q}$ and $\mathcal{P}_{N}$ is a $2^{N}\times2^{N}$ permutation matrix taking the computational basis to a chosen excitation basis. There are many choices of excitation basis, as one can always reorder computational basis states within a given $q$-block to obtain a new basis.

Next, one can leverage the translation symmetry of the models. For homogeneous couplings and the two topologies considered here, the Hamiltonian is invariant under relabelings like $i\rightarrow i-1$. That is, the translation operator over $N$ spins, $T_{N}$, satisfies $[T_{N},H_{\rm XXZ}]=0$. Thus the next stage in block-diagonalization for both models is achieved by using a Fourier basis, with blocks labelled by $a$:
\begin{equation}
    \Gamma_{N}\mathcal{P}_{N}\left(H_{\rm XXZ}\right)\mathcal{P}_{N}^{\dag}\Gamma_{N}^{\dag}=\bigoplus_{q=0}^{N}\bigoplus_{a=0}^{N-1}H^{q,a}_{\rm XXZ}\,,
\end{equation}
Where $\Gamma_{N}$ is the transformation that diagonalizes the translation operator $T_{N}$. See the Appendix for additional details and examples. 

However, for $N>3$, even these symmetries are not enough to achieve exact diagonalization (see, for example, Eq.(\ref{eq:4Qextradegenerate}) in the Appendix). The additional degeneracy in the eigenstates of $T_{N}$ increases with $N$, and so eigenstates of $H_{\rm XXZ}$ are superpositions of these Fourier modes. That is, if eigenstates of the Hamiltonian are labelled by a complete set of quantum numbers \{$q$,$a$,$l$\}, and the index $k$ runs over the Fourier modes in each block $a$, then
\begin{equation}
\begin{split} 
|E^{a}_{q};l\rangle&=\sum_{k=k_{\rm min}}^{k_{\rm max}} C_{q}^{kl}|\mathcal{F}^{a}_{q};k\rangle 
 \,.
\end{split}
\end{equation}
Whenever $k_{\rm max}=k_{\rm min}$, the stationary states are just Fourier modes i.e. 
$|E^{a}_{q};k_{\rm min}\rangle=|\mathcal{F}^{a}_{q};k_{\rm min}\rangle$. For small $(k_{\rm max}-k_{\rm min})$, the time evolution may be computed analytically. That is the procedure we will follow in the examples below, but the maximum degree of degeneracy amongst Fourier modes grows linearly with $N$, and it becomes more difficult to perform the diagonalization.  

To make this a bit more explicit, consider the $q=1$ block for an XXZ-network with $N$ qubits. The block has dimension $N$ and $T_{N}$ is non-degenerate within it, so diagonalization of $H^{q=1}_{\rm XXZ}$ is achieved through an $N$-dimensional discrete Fourier transformation. Explicitly the eigenstates are given as discrete plane waves
\begin{equation}
|\mathcal{F}^{a}_{1};0\rangle \equiv\frac{1}{\sqrt{N}}\sum_{k=0}^{N-1}e^{i k\theta_{a}}\sigma^{+}_{k}|1...1\rangle\,,
\end{equation}
where $\theta_{a}=\frac{2 \pi a}{N}$ and $\sigma^{+}_{k}$ flips the $k$th spin from 1 to 0 and acts as the identity every where else. The $q=N-1$ block has essentially identical eigenstates, expect with the replacement
\begin{equation}
\sigma^{+}_{k}|1...1\rangle\rightarrow \sigma^{-}_{k}|0...0\rangle \,,
\end{equation}
with $\sigma^{-}_{k}=(\sigma^{+}_{k})^{\dag}$.
For other values of $q$, the translation operator becomes more degenerate. In these cases an appropriately sized discrete Fourier transformation must be performed to reach the next step of block-diagonalization. 

In the remainder of this subsection, we go through the computation of the dynamical maps for three qubits in detail. Then, in the rest of the section, we report map parameters and average maps for $N>3$, leaving details for larger N to Appendix A. 

%%%%%%%%%%%%%%%%%%%%%%%%%%%%%%%%%%%%%%%%%%
%%%%%%%%%%%%%%%%%%%%%%%%%%%%%%%%%%%%%%%%%%
\subsubsection{Reduced dynamics of XXZ-networks}
%%%%%%%%%%%%%%%%%%%%%%%%%%%%%%%%%%%%%%%%%%
%%%%%%%%%%%%%%%%%%%%%%%%%%%%%%%%%%%%%%%%%%
We shall now demonstrate how we compute the ensemble of phase-covariant maps and the average map, and characterize the fluctuations, for the example of a 3-qubit network. In this case, the ring (nearest-neighbor interactions and no boundary spins) and completely connected (interactions between every pair of spins) connectivities are equivalent. We begin by diagonalizing $H_{\rm XXZ}$, going first to the canonically ordered excitation basis
\begin{equation}
\mathcal{P}_{3}H_{\rm XXZ}\mathcal{P}_{3}^{\dag}= H^{0}_{\rm XXZ}\oplus H^{1}_{\rm XXZ}\oplus H^{2}_{\rm XXZ}\oplus H^{3}_{\rm XXZ} \,.
\end{equation} 
This example is especially simple as the only interacting blocks, $H^{1}_{\rm XXZ}$ and $H^{2}_{\rm XXZ}$, are both diagonalized through discrete Fourier transformation. The other blocks are non-interacting, and consist of the computational basis states $|1... 1\rangle$ ($q$=0) and $|0...0\rangle$ ($q$=$N$).
Using the exact diagonalization of the Hamiltonian, we determine explicit expression for the parameters appearing in the phase-covariant dynamical maps when the initial state is of the form Eq.(\ref{eq:IC}). We find the dynamical map describing the evolution of the $i$th spin is given in terms
\begin{equation}
\begin{split}
    &\lambda_{1,i}(t,3)=\sqrt{\alpha_{i}^{2}(t,3)+\beta_{i}^{2}(t,3)}  \\
    &\theta_{i}(t,3)=2ht+\phi_{i}(t,3) \\
    &\lambda_{3,i}(t,3) = \frac{1}{9}\big(5+4\cos{3 J_{\perp}t}\big)\\
    &\tau_{3,i}(t,3) = \frac{2}{9}(\sum_{k \neq i}z_{k})\big(1-\cos{3 J_{\perp}t}\big) \,.
\end{split}
\end{equation}
Where $z_{k}$ are the Bloch components of the remaining qubits (spins) in the network defined in Eq.(\ref{eq:IC}).
To simplify the notation we have made the following definitions
\begin{equation}
    \begin{split}
        \alpha_{i}(t,3)&=\frac{1}{18}\Big(\big(1-\prod_{k\neq i}z_{k}\big)\big(7+2\cos{3J_{\perp}t}\big)\\ &+\big(1+\prod_{k\neq i}z_{k}\big)\big(3\cos{(2J_{\parallel}-2J_{\perp})t}+6\cos{(J_{\parallel}+2J_{\perp})t}\big)\Big) \\
        \beta_{i}(t,3)&=\frac{1}{18}\big(\sum_{k \neq i}z_{k}\big)\big(3\sin{(2J_{\parallel}-2J_{\perp})t}+6\sin{(J_{\parallel}+2J_{\perp})t}\big) \\ 
        \phi_{i}(t,3)&= {\rm tan}^{-1}\left(\frac{\beta_{i}(t,3)
        }{\alpha_{i}(t,3)}\right)\,.
    \end{split}
\end{equation}
Notice that the strength of the external magnetic field, $h$ only enters into the rotation angle $\theta$.

There are two relevant time scales in these maps, determined by $J_{\parallel}$ and $ J_{\perp}$. For simplicity, we will illustrate results in Figure 1 for $J_{\parallel}= J_{\perp}$, and define a single time-scale 
\begin{equation}
\label{eq:timesc}
  t_{J}=\frac{2\pi}{J_{\perp}}\,.
\end{equation}
While a simplifying assumption, we do not expect this reduction to drastically change the primary aspects of our results for the average phase-covariant maps computed below. For example, the late-time average values of both $\lambda_{3}$ and $\tau_{3}$ are independent of the ratio $\frac{J_{\perp}}{J_{\parallel}}$. Further, if we avoid certain commensurate values of Hamiltonian parameters, the steady-state values of $\lambda_{1}$ are also independent of the coupling ratio. 

\begin{figure}[H]
\centering
\begin{subfigure}{.45\linewidth}
  \centering
\includegraphics[width=0.95\linewidth]{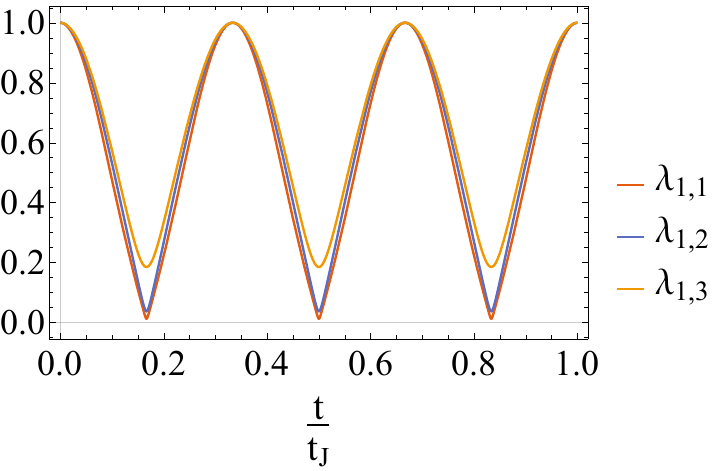} 
\end{subfigure} 
\begin{subfigure}{.45\linewidth}
  \centering
\includegraphics[width=0.95\linewidth]{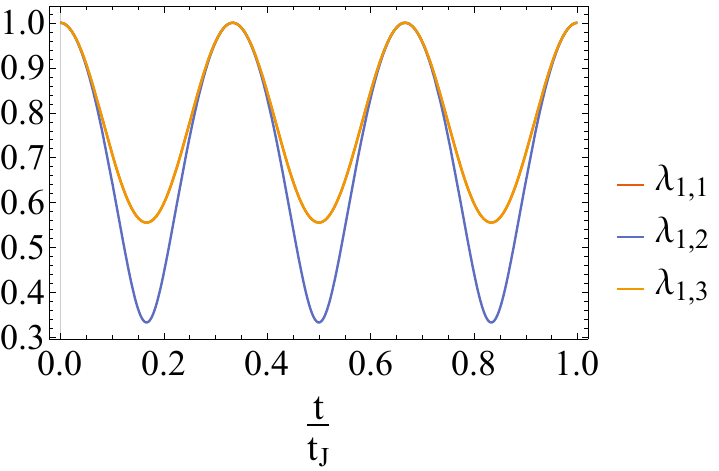}
\end{subfigure}
\begin{subfigure}{.45\linewidth}
  \centering
\includegraphics[width=0.95\linewidth]{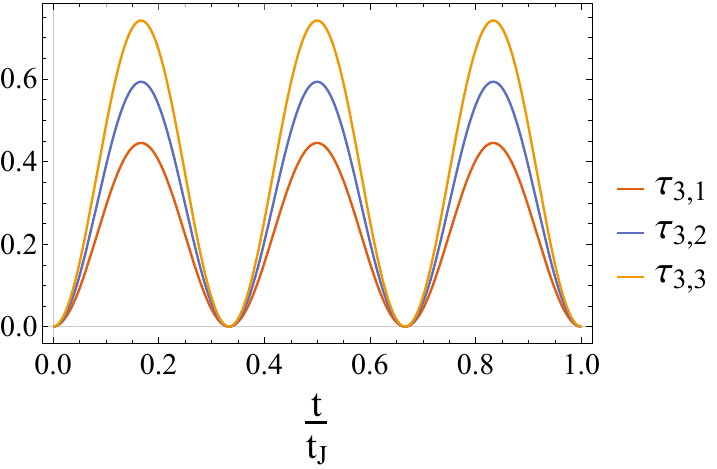} 
\caption{Hierarchy}
\end{subfigure} 
\begin{subfigure}{.45\linewidth}
  \centering
\includegraphics[width=0.95\linewidth]{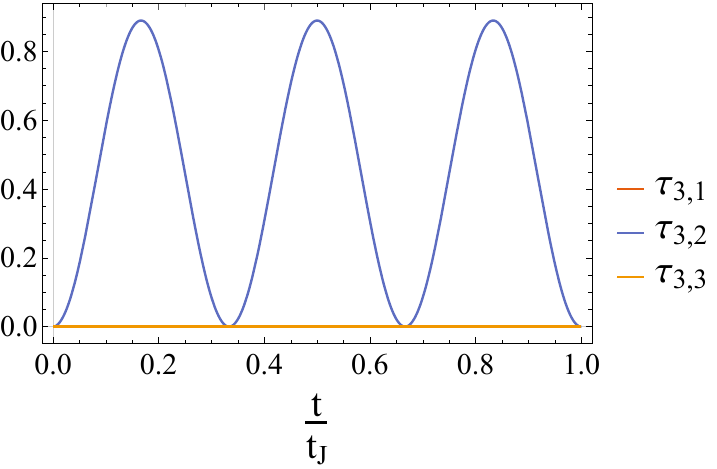}
 \caption{N\'{e}el}
\end{subfigure}
\caption{Examples of ensembles of two of the dynamical map functions, $\lambda_{1,i}(t)$ and $\tau_{3,i}(t)$. Each panel shows results for all spins ($i=1,2,3$) in the network. Plots on the left are for initial state $z_{1}$=1, $z_{2}=\frac{1}{3}$ and $z_{3}=\frac{2}{3}$ (hierarchy), while those on the right had $z_{1}=z_{3}=-z_{2}=1$ (N\'{e}el). Since in the N\'{e}el initial state the $1^{st}$ and $3^{rd}$ qubits have the same initial state, and the same environmental initial state, they have identical dynamical maps and so two of the lines coincide on the left panels. For all cases we take $J_{\parallel}=J_{\perp}$, which is used to define the time scale $t_{J}=\frac{2\pi}{J_{\perp}}$. \label{fig:l3}}
\end{figure}
We now turn to the computation of the steady-channel, $\overline{\Lambda}_{3}^{\infty}$ and the normalized fluctuations. For generic values of $h$ (i.e. not commensurate with $J_{\perp}$ or $J_{\parallel}$), $\Lambda_{xx}(t)$, the components of the dynamical map matrix (with labels matching the Pauli basis vector for the density matrix) $\Lambda_{xy}(t)$, $\Lambda_{yx}(t)$, and $\Lambda_{yy}(t)$, vanish under long-time averaging. Therefore we assume that $h$ is generic, precluding this situation. Choosing to define the time average of $\lambda_{1}(t)$ as
\begin{equation}
    \overline{\lambda}_{1}(t)=\sqrt{\overline{\Lambda}^{2}_{xx}(t)+\overline{\Lambda}^{2}_{xy}(t)}
\end{equation}
we then find
\begin{equation}
\begin{split}
    &\overline{\lambda}_{1}(t=\infty,N=3)=0 \,,\\
    &\overline{\lambda}_{3}(t=\infty,N=3)=\frac{5}{9}\,, \\
    &\overline{\tau}_{3}(t=\infty,N=3)=\frac{4}{9}\frac{\mathfrak{e}_{1}(z_{i})}{3}\,.
\end{split}
\end{equation}
Anticipating results for larger $N$, we express $\overline{\tau}_{3}^{\infty}$ in terms of elementary symmetric polynomials in the initial state $\mathfrak{e}_{k}(z_{i})$. These polynomials are defined for any $N$ by the equation
\begin{equation}
    \prod_{i=1}^{N}(z+z_{i})=\mathfrak{e}_{0}(z_{i})z^{N}+\mathfrak{e}
    _{1}(z_{i})z^{N-1}+...+\mathfrak{e}_{N}(z_{i})\,.
\end{equation}
For example we have  $\mathfrak{e}_{0}(z_{i})=1$, $\mathfrak{e}_{1}(z_{i})=\sum_{k}z_{k}$, and so on.  
Additional examples and related applications of the polynomials can be found in \cite{macdonald1998symmetric}. 

Before we move on to the next topic, it is useful to point out a few features of the dynamical map ensembles for the various $N$ we consider. 
\begin{figure}[H]
\centering
\begin{subfigure}{.45\linewidth}
  \centering
\includegraphics[width=0.95\linewidth]{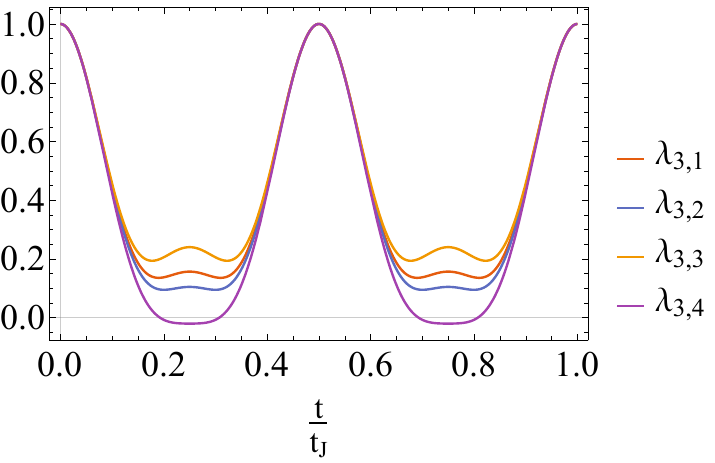} 
\caption{$\lambda_{3}$ (4 spins)}
\end{subfigure} 
\begin{subfigure}{.45\linewidth}
  \centering
\includegraphics[width=0.95\linewidth]{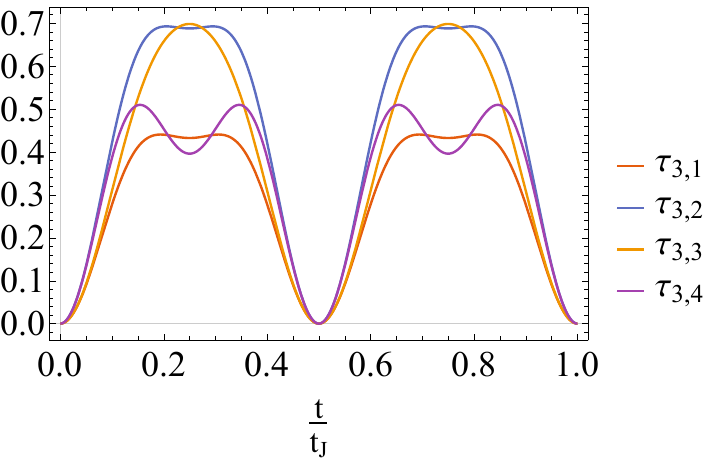} 
 \caption{$\tau_{3}$ (4 spins)}
\end{subfigure}
\begin{subfigure}{.45\linewidth}
  \centering
\includegraphics[width=0.9\linewidth]{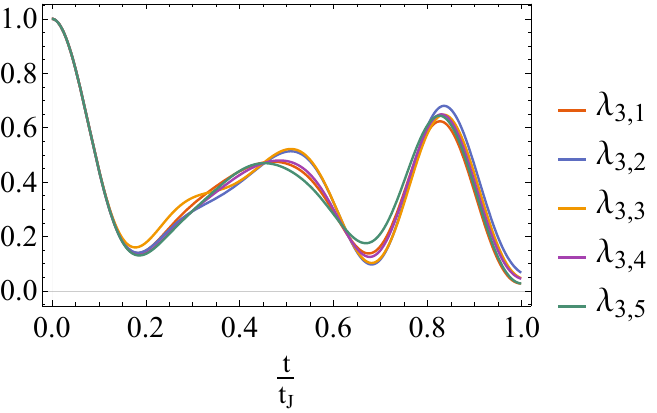} 
\caption{$\lambda_{3}$ (5 spins)}
\end{subfigure} 
\begin{subfigure}{.45\linewidth}
  \centering
\includegraphics[width=0.9\linewidth]{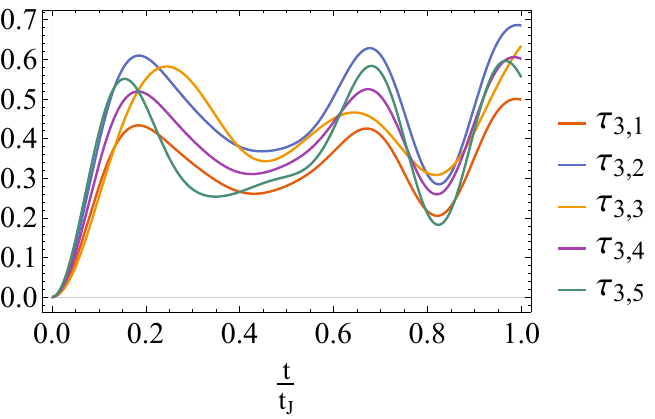}
 \caption{$\tau_{3}$ (5 spins)}
\end{subfigure}
\caption{The ensembles of non-unitary parameters are plotted for the 4 spin (top row) and 5 spin (bottom row) XXX model with ring connectivity. The upper plots are generated using the the initial state with $z_{1}=1$, $z_{2}=\frac{1}{4}$, $z_{3}=\frac{2}{4}$, and $z_{4}=\frac{3}{4}$. The lower plots are generated using the initial condition $z_{1}=1$, $z_{2}=\frac{1}{5}$, $z_{3}=\frac{2}{5}$, $z_{4}=\frac{3}{5}$, and $z_{5}=\frac{4}{5}$.}
\label{fig:t4}
\end{figure}
Figure 2 emphasizes the time-nonlocal nature or non-invertibility of our ensembles, following from the existence of zeros in the $\lambda_{3}(t)$ plots. The top left panel shows that while not every individual map is time-local for the $N=4$ hierarchy state, the average dynamics will be. While not plotted here, the N\'{e}el serves as a maximal example of locality breaking, as even the average dynamical map is time-non-local.

We may also use these plots to draw a point of distinction between the two topologies of networks. In the completely connected case $\lambda_{3}(t)$ and $\tau_{3}(t)$ are always periodic over some multiple of $t_{J}$, while in the case of the XXX-ring the dynamics need not be periodic, as can be seen from the lower panel of Figure 2. Of course if $N$ is a perfect square then periodicity occurs, as can be seen for the case of $N=4$. Equations (\ref{eq:l34qubitting}) and (\ref{eq:l35qubitring}) in the Appendix makes this point clear, showing that for the XXX-ring $\sqrt{N}J_{\perp}$ appears in the various frequencies comprising the dynamical map components.  
%%%%%%%%%%%%%%%%%%%%%%%%%%%%%%%%%%%%%%%%%%
%%%%%%%%%%%%%%%%%%%%%%%%%%%%%%%%%%%%%%%%%%
\subsection{Long-time averaged channels}
%%%%%%%%%%%%%%%%%%%%%%%%%%%%%%%%%%%%%%%%%%
%%%%%%%%%%%%%%%%%%%%%%%%%%%%%%%%%%%%%%%%%%
The individual maps and steady-channels for $N>3$ require a similar, but longer computation. We also discuss features of the reduced dynamics found for the computed ensembles, i.e. whether they are time-local for certain initial states and other average properties. The dynamical map components as a function of time for the completely connected network are found in Eq.(\ref{eq:CC})-Eq.(\ref{eq:CC8}). And the dynamical map components as a function of time for the ring connected network are found in Eq.(\ref{eq:l34qubitting})-Eq.(\ref{eq:RC}).

%%%%%%%%%%%%%%%%%%%%%%%%%%%%%%%%%%%%%%%%%%
\subsubsection{Complete connectivity} 
%%%%%%%%%%%%%%%%%%%%%%%%%%%%%%%%%%%%%%%%%%
Turning our attention to the completely-connected XXZ-network we compute the late-time network averaged quantum channels. For $N=4$ we find 
\begin{equation}
\label{eq:ccQ4}
    \begin{split}
        \overline{\lambda}^{\infty}_{1}(N=4)&=0 \\
        \overline{\lambda}^{\infty}_{3}(N=4)&=\frac{7}{16}\frac{\mathfrak{e}_{0}(z_{i})}{1}+\frac{3}{16}\frac{\mathfrak{e}_{2}(z_{i})}{6} \\   \overline{\tau}^{\infty}_{3}(N=4)&=\frac{9}{16}\frac{\mathfrak{e}_{1}(z_{i})}{4}-\frac{3}{16}\frac{\mathfrak{e}_{3}(z_{i})}{4} \, .
    \end{split}
\end{equation}
 For $N=5$ we find
\begin{equation}
\label{eq:ccQ5}
    \begin{split}
        \overline{\lambda}^{\infty}_{1}(N=5)&=0 \\
        \overline{\lambda}^{\infty}_{3}(N=5)&=\frac{7}{15}\frac{\mathfrak{e}_{0}(z_{i})}{1}+\frac{16}{75}\frac{\mathfrak{e}_{2}(z_{i})}{10}\\
        \overline{\tau}^{\infty}_{3}(N=5)&=\frac{8}{15}\frac{\mathfrak{e}_{1}(z_{i})}{5}-\frac{16}{75}\frac{\mathfrak{e}_{3}(z_{i})}{10}\,. 
    \end{split}
\end{equation}
and for $N=6$ we find
 \begin{equation}
    \begin{split}
        &\overline{\lambda}^{\infty}_{1}(N=6)=0 \\ &\overline{\lambda}^{\infty}_{3}(N=6)=\frac{59}{144}\frac{\mathfrak{e}_{0}(z_{i})}{1}+\frac{5}{12}\frac{\mathfrak{e}_{2}(z_{i})}{15}-\frac{5}{48}\frac{\mathfrak{e}_{4}(z_{i})}{15} \\
        &\overline{\tau}^{\infty}_{3}(N=6)=\frac{85}{144}\frac{\mathfrak{e}_{1}(z_{i})}{6}-\frac{5}{12}\frac{\mathfrak{e}_{3}(z_{i})}{20}+\frac{5}{48}\frac{\mathfrak{e}_{5}(z_{i})}{6}\,. 
    \end{split}
\end{equation}
The relationship appearing between the first coefficient in the expansion of $\overline{\lambda}^{\infty}_{3}$ and the first in the expansion of $\overline{\tau}^{\infty}_{3}$ is due to the following relations,  
\begin{equation}
\label{eq:fixedpoint}
\overline{\tau}^{\infty}_{3}\Big|_{z_{i}=\pm 1}\pm \overline{\lambda}^{\infty}_{3}\Big|_{z_{i}=\pm 1}=\pm 1 \,.
\end{equation}
which are derived for $z_{i}=\pm 1$ for values of $i$. These two initial state are fixed point of the unitary evolution resulting in Eq.(\ref{eq:fixedpoint}).

Another property of these long time averages, and in fact of the dynamical map components in general, is that $\lambda_{3}$ is an even function of $\{z_{i}\}$ while $\tau_{3}$ is an odd function. This is a consequence of the symmetry generated by  
\begin{equation}
    \mathcal{P}=\prod_{i=1}^{N}Z_{i}.
\end{equation}
which is a symmetry for both models considered, therefore $\lambda_{3}(t)$ and $\tau_{3}(t)$ have definite values of $\mathcal{P}$ for the ring connectivity as well.

 To conclude the discussion, notice that the average map depends only on the initial state and symmetry of the Hamiltonian. We note that in this case the long time averages are invariant under any permutation of the initial $z_{i}$. Comparing the fully connected case to the ring, shown next, illustrates this dependence. In addition, notice that the invariant state of the average map is not the average of the initial states, but any explicit dependence on coupling strength is gone. As expected from the individual maps at any time, there is also no dependence on magnetization. In the case of non-phase covariant maps, there will be terms depending on $h$, and, just as in the case of $\lambda_{1}$, special values of the coupling lead to different long time average values.   

%%%%%%%%%%%%%%%%%%%%%%%%%%%%%%%%%%%%%%%%%%
\subsubsection{Ring connectivity}
%%%%%%%%%%%%%%%%%%%%%%%%%%%%%%%%%%%%%%%%%%
 Computing the late-time and ring-averaged quantities we find for $N=4$
\begin{equation}
\label{eq:ring4}
    \begin{split}
    \overline{\lambda}^{\infty}_{1}(N=4)&=0 \\
    \overline{\lambda}^{\infty}_{3}(N=4)&=\frac{7}{16}+\frac{3}{16}\frac{\mathfrak{e}_{2}(z_{i})-4z_{1}z_{3}-4z_{2}z_{4}}{6}\\  \overline{\tau}^{\infty}_{3}(N=4)&=\frac{9}{16}\frac{\mathfrak{e}_{1}(z_{i})}{4}+\frac{1}{16}\frac{\mathfrak{e}_{3}(z_{i})}{4} \,,
    \end{split}
\end{equation}
and for $N=5$
\begin{equation}
\label{eq:ring5}
    \begin{split}
    \overline{\lambda}^{\infty}_{1}(N=5)&=0 \\
    \overline{\lambda}^{\infty}_{3}(N=5)&=\frac{71}{225}+\frac{2}{45}\frac{\mathfrak{e}_{2}(z_{i})}{10}\\
        \overline{\tau}^{\infty}_{3}(N=5)&=\frac{154}{225}\frac{\mathfrak{e}_{1}(z_{i})}{5}-\frac{2}{45}\frac{\mathfrak{e}_{3}(z_{i})}{10} \,.
    \end{split}
\end{equation}
Comparing to the results for the completely connected network, notice that the symmetry structure of the Hamiltonian is embedded into the long-time averages. For example taking $N=4$, the long-time average is only invariant under cyclic permutations of the initial state while in the fully connected case the invariance was with respect to the full permutation group. However, for $N=5$ the full permutation symmetry is recovered in the infinity time limit. But we do not expect this to be a generic feature for larger $N$, and in principle only cyclic polynomials of the initial state should appear. As before, the invariant state of the average map does not correspond to the average initial state.

\subsection{Fluctuations about the average channel}
 We now consider physical data to constrain the variance we expect in the distributions at late-times. We do this by considering the maximum distance from the finite-time average to the steady value. We define the fluctuations at finite-time about these average quantities, estimated by the normalized fluctuations
\begin{equation}
    \begin{split}
        \Delta\lambda_{i}&\equiv \frac{\overline{\lambda}_{i}(t)-\overline{\lambda}_{i}^{\infty}}{\overline{\lambda}_{i}^{\infty}}  \\
         \Delta\tau_{3}&\equiv \frac{\overline{\tau}_{3}(t)-\overline{\tau}_{3}^{\infty}}{\overline{\tau}_{3}^{\infty}} \, .
    \end{split}
\end{equation}
These quantities are used to calibrate the second moments of the phase-covariant measures developed in Section 5. The normalized fluctuations are generally initial-state dependent, but to determine an approximate size of the fluctuations, we consider a class of initial states with essentially initial state independent fluctuations. To see that such a class of states exists, consider the completely connected network and assume that each qubit satisfies $0\leq z_{i}\leq \mathcal{Z}<1$. Since
\begin{equation}
    \frac{\mathfrak{e}_{k}(z_{i})}{{N\choose k}}\leq \mathcal{Z}^{k}\,,
\end{equation}
then for small enough $\mathcal{Z}$ only the first terms in $\overline{\lambda}_{3}(t)$ and $\overline{\tau}_{3}(t)$ are relevant. In such a class the fluctuations about the long time average will be initial state independent. 

Figures 3 and 4 show the fluctuations about the long time averages for the completely and ring connected networks, using only the lowest order terms in the initial state from $\overline{\lambda}_{3}(t)$ and $\overline{\tau}_{3}(t)$. We confirm in Section 5 that these approximations provide a reasonable estimate of the fluctuation size, even taking a generic initial state. 
\begin{figure}[H]
\centering
\begin{subfigure}{.45\linewidth}
  \centering
\includegraphics[width=0.95\linewidth]{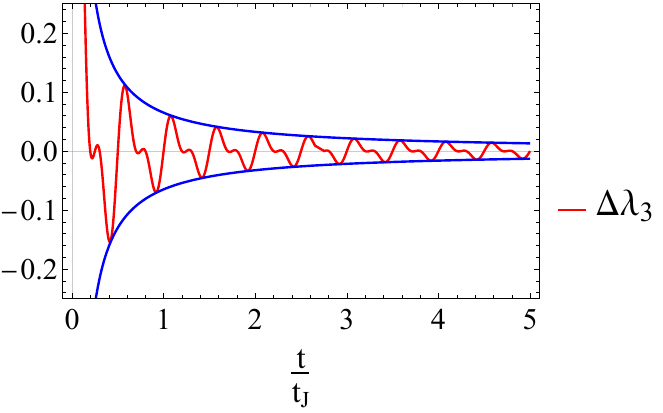} 
\caption{$C_{\lambda}\approx0.26$\,\,(4 spins)}
\end{subfigure} 
\begin{subfigure}{.45\linewidth}
  \centering
\includegraphics[width=0.95\linewidth]{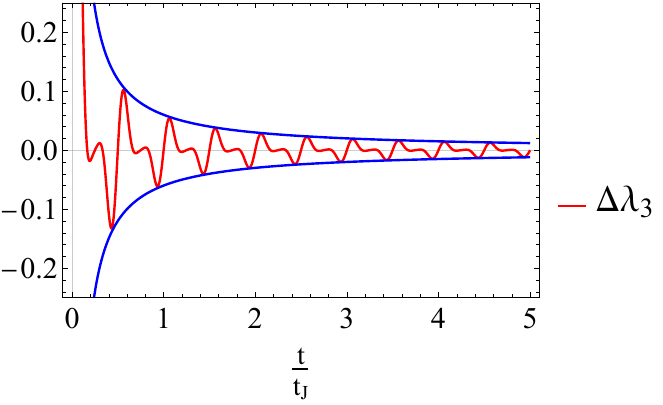} 
\caption{$C_{\lambda}\approx0.36$\,\,(6 spins)}
\end{subfigure}
\begin{subfigure}{.45\linewidth}
  \centering
\includegraphics[width=0.95\linewidth]{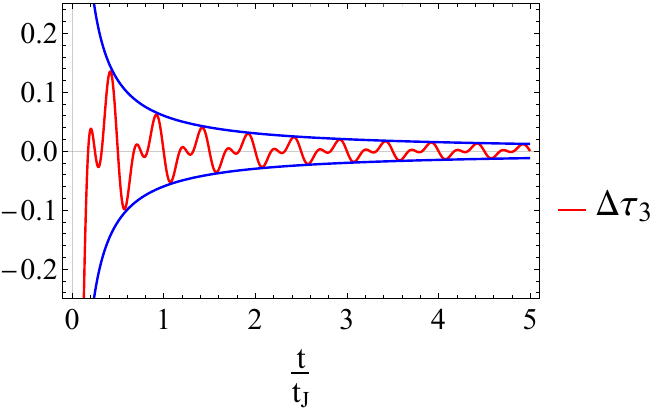} 
\caption{$C_{\tau}\approx 0.24$\,\,(4 spins)}
\end{subfigure} 
\begin{subfigure}{.45\linewidth}
  \centering
\includegraphics[width=0.95\linewidth]{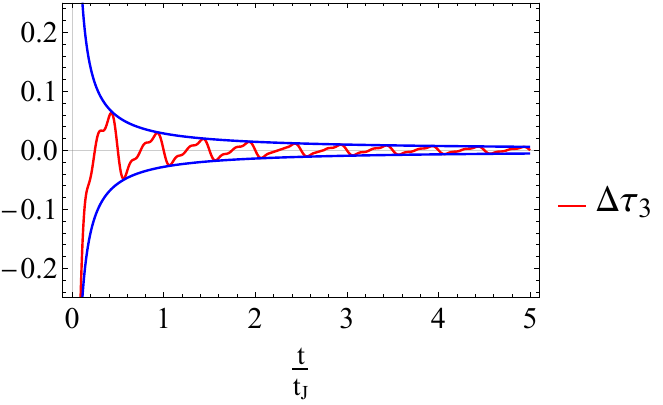} \caption{$C_{\tau}\approx 0.17$\,\,(6 spins)}
\end{subfigure}
\caption{The normalized fluctuations of $\overline{\lambda}_{3}(t)$ and $\overline{\tau}_{3}(t)$  are plotted for the completely-connected XXZ network, considering only the lowest order terms in the initial state. The bounding hyperbolas (in blue) are of the form $\pm\frac{C_{\zeta}t_{J}}{Nt}$, with the approximate value of $C_{\zeta}$ listed beneath each graph.  The bounding hyperbolas provide an estimate for the size of the late-time fluctuations, as in Eq.(\ref{eq:DeltaChi}). }
\label{fig:ccfluc}
\end{figure}
These plots demonstrates that the late-time fluctuations in any of the non-unitary parameters (let $\zeta$ label any member of that set) are bounded by
\begin{equation}
\label{eq:DeltaChi}
    |\Delta\zeta|\leq  \frac{C_{\zeta}t_{J}}{Nt}\,,
\end{equation}
where $t_{J}=\frac{2\pi}{J_{\perp}}$ is the dynamical timescale defined in Eq.(\ref{eq:timesc}), $C_{\zeta}$ is O(1), and N is the number of spins. The completely connected network, for a fixed $N$, approaches the steady channel more quickly than in the case of the ring. Comparing the $N=6$ completely connected network and the $N=5$ ring network, we see that the fluctuations persist at least four times longer in the case of the ring. Note that the scale of the vertical axis of all plots are identical.

\begin{figure}[H] 
\centering
\begin{subfigure}{.45\linewidth}
  \centering
\includegraphics[width=0.95\linewidth]{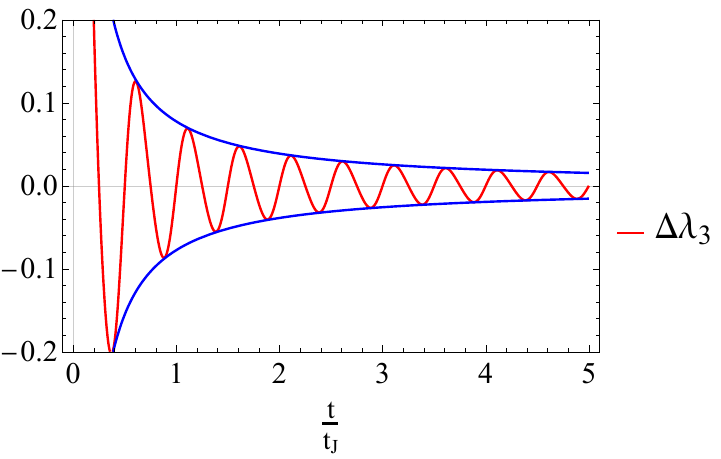} 
\caption{$C_{\lambda}\approx0.31$\,\,(4 spins)}
\end{subfigure} 
\begin{subfigure}{.45\linewidth}
  \centering
\includegraphics[width=0.95\linewidth]{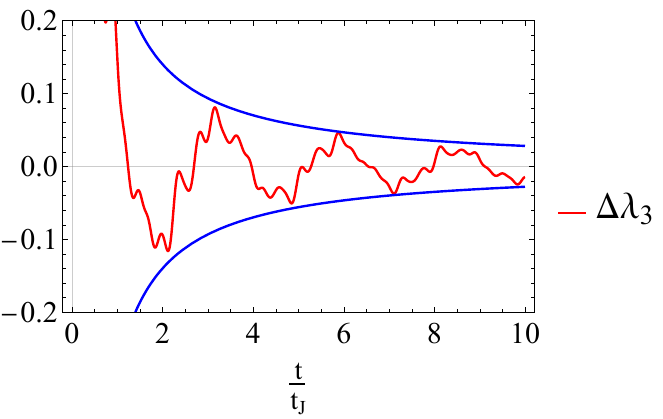}
 \caption{$C_{\lambda}\approx1.40$\,\,(5 spins)}
\end{subfigure}
\centering
\begin{subfigure}{.45\linewidth}
  \centering
\includegraphics[width=0.95\linewidth]{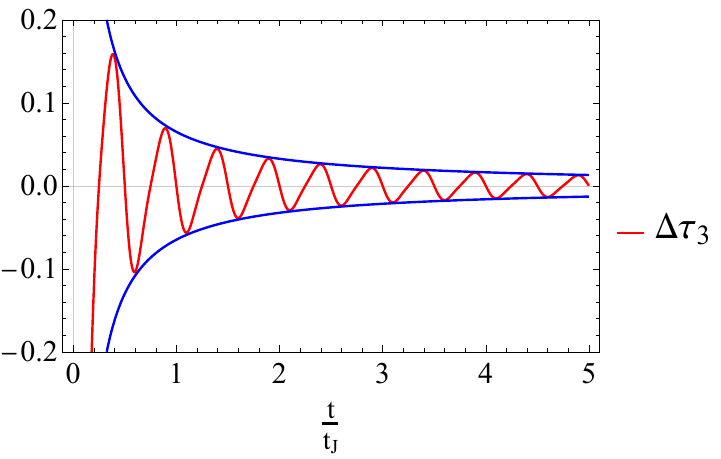} 
\caption{$C_{\tau}\approx0.26$\,\,(4 spins)}
\end{subfigure} 
\begin{subfigure}{.45\linewidth}
  \centering
\includegraphics[width=0.95\linewidth]{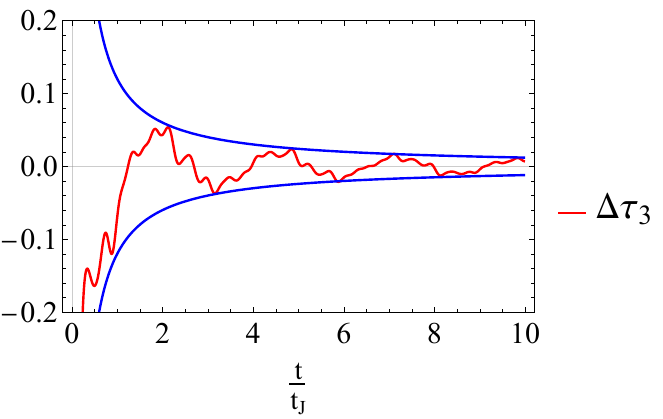}
 \caption{$C_{\tau}\approx 0.59$\,\,(5 spins)}
\end{subfigure}
\caption{The normalized fluctuations of $\overline{\lambda}_{3}(t)$ and $\overline{\tau}_{3}(t)$  are plotted for the XXZ ring, considering only the lowest order terms in the initial state. The bounding hyperbolas (in blue) are of the form $\pm\frac{C_{\zeta}t_{J}}{Nt}$, with the approximate value of $C_{\zeta}$ listed beneath each graph. The bounding hyperbolas provide an estimate for the size of the late-time fluctuations, as in Eq.(\ref{eq:DeltaChi}).}
\label{fig:ringfluc}
\end{figure}

%%%%%%%%%%%%%%%%%%%%%%%%%%%%%%%%%%%%%%%
%%%%%%%%%%%%%%%%%%%%%%%%%%%%%%%%%%%%%%%
\section{Physical relevance of the long-time averaged channel}
\label{sec:JustifyLambdaBar}
\setcounter{equation}{0}
%%%%%%%%%%%%%%%%%%%%%%%%%%%%%%%%%%%%%%%
%%%%%%%%%%%%%%%%%%%%%%%%%%%%%%%%%%%%%%%
 Consider a qubit network that consists of $N_{\rm CL}$ clusters, each containing $N$ qubits, where we assume  $N_{\rm CL}\gg N$. For simplicity, each cluster is assumed to have the same connectivity. The spin-averaged dynamical map for the $I$th cluster is 
 \begin{equation}
 \label{eq:BenchM}
 \begin{split}
\langle\Lambda_{\rm I}(t)\rangle&=\frac{1}{N}\sum_{i=1}^{N
    }\Lambda_{\rm I_{i}}(t) \,.  \\
\end{split}
 \end{equation}
 Suppose at some very early time there are no interactions and the state of the network is entirely uncorrelated. Interactions within each individual cluster are turned on via a quenching mechanism, but that not all clusters are quenched at the same time. Let the time-scale for over which the cluster quenching is (uniformly) staggered be longer than that set by the interaction time-scale within each cluster. The time-dependent Hamiltonians considered (and initial states) are such that the conditions for phase covariance are satisfied for each cluster. For example, this scenario can be modeled by a Hamiltonian of the form
 \begin{equation}
   H(t)= \sum_{\rm I=1}^{N_{\rm CL}}\Big( h_{\rm I}\sum_{i=1}^{N} Z_{\rm I_{i}}+J\theta\big(t-t_{\rm I}\big)\sum_{i,j}\big(X_{\rm I_{i}}X_{\rm I_{j}}+Y_{\rm I_{i}}Y_{\rm I_{j}}+Z_{\rm I_{i}}Z_{\rm I_{j}}\big)\Big)
 \end{equation}
 where $\theta(t)$ is the Heaviside step function, and $t_{\rm I}$ are the various quench times associated with the clusters. If the timing of the quench varies over a time-scale longer than that set by $J$, and the large number of clusters ensures that all oscillation scales of the single-cluster Hamiltonian are smoothly sampled, then the clusters will be at varying points in their evolution and the average map at late times will be given by
 \begin{equation}
\label{eq:networkcond}
    \lim_{N_{\rm CL}\rightarrow \infty }\frac{1}{N_{\rm CL}}\sum_{\rm I=1}^{N_{\rm CL}}\langle\Lambda_{\rm I}(t)\rangle  =\frac{1}{t}\int_{0}^{t}dt'\langle\Lambda_{\rm K}(t')\rangle=\overline{\Lambda}_{N}(t)\,.
\end{equation}
In fact, as long as there are many clusters with any given initial state, which experience the quench at offset times, one can imagine a further average over steady channels with different initial states to match the full complexity of the initial system of size $\gg N$. That is, one could average over the initial states appearing in the steady channels.

Under such a quenching scenario, even for relatively small clusters, we obtain average dynamical maps that approach a steady-channel, denoted as $\overline{\Lambda}_{N}^{\infty}$, in the limit $t \rightarrow \infty$. We are then able to construct distributions of channels that have the steady channel as their average. By studying the approach to the steady channel, we able to constrain the second moment of the distributions.

%%%%%%%%%%%%%%%%%%%%%%%%%%%%%%%%%%%%%%%
%%%%%%%%%%%%%%%%%%%%%%%%%%%%%%%%%%%%%%%
\section{Phase-covariance on average from disordered Hamiltonians} 
\label{sec:QCensembles}
\setcounter{equation}{0}
%%%%%%%%%%%%%%%%%%%%%%%%%%%%%%%%%%%%%%%
%%%%%%%%%%%%%%%%%%%%%%%%%%%%%%%%%%%%%%%
While in the staggered quench example above we imposed that each cluster had the same Hamiltonian and an initial state from the phase-covariant set, the regime of applicability of the phase-covariant steady channel is actually larger. In this section we show that constrained Hamiltonian disorder, or noise, can be rather naturally consistent with phase-covariant, average reduced dynamics. We show this by considering ensembles of qubit pairs, which serve as a simple model of the paradigm set forth at the end of the previous section.

%%%%%%%%%%%%%%%%%%%%%%%%%%%%%%%%%%%%%%%
%%%%%%%%%%%%%%%%%%%%%%%%%%%%%%%%%%%%%%%
\subsection{An ensemble with non-phase-covariant elements}
%%%%%%%%%%%%%%%%%%%%%%%%%%%%%%%%%%%%%%%
%%%%%%%%%%%%%%%%%%%%%%%%%%%%%%%%%%%%%%%
 To begin we start with a physical setting where distributions of maps that are not necessarily phase-covariant appear. We consider a Hamiltonian that governs the time evolution of $N_{\rm CL}$ pairs of spins, where each pair evolves independently, and conserves the total excitation number $Q=\sum Z_{i}$. This can be compactly written as
\begin{equation}
\begin{split}
H&=\sum_{i=1}^{2N_{\rm CL}} h_i Z_{i}+\sum_{i=1}^{2N_{\rm CL}}\frac{J_{i(i+1)}}{2}\left[\frac{1+(-1)^{i+1}}{2}\right]\left[X_{i}X_{i+1}+Y_{i}Y_{i+1}\right] \\ 
&=\sum_{i=1}^{2N_{\rm CL}} \frac{1+(-1)^{i+1}}{2}H^{i,i+1}_{\rm XX}=H^{1,2}_{\rm XX}+H^{3,4}_{\rm XX}+...+H^{2N-1,2N_{\rm CL}}_{\rm XX}\,.
\end{split}
\end{equation}
The $2N_{\rm CL}$ qubits are taken to be in a completely uncorrelated initial state. For each pair, we allow this example to break both phase-covariant conditions from Eq.({\ref{eq:PCcond}}). First, allowing $h_i\neq h_j$ means that the interaction term will not commmute with the free Hamiltonian. In addition, we do not impose any restriction on the initial state. 

As the Hamiltonian is block diagonal in the qubit pairs, we can determine both the 1-spin and 2-spin dynamical maps obtained for a single pair. The 2-qubit maps describe unitary evolution, and this example may be viewed as an generalization of the noisy qubit models proposed in \cite{Kropf:2016} to a larger system size. However, at the single-spin level the evolution is non-unitary and richer than what can be found with noisy Hamiltonians alone.

To write the single-spin dynamical map, it is convenient to decompose the unitary evolution in terms of 
\begin{equation}
\mathcal{U}^{ij}_{lk}\equiv\frac{1}{4}{\rm tr}\left[\big(\sigma^{i}_{\rm 1}\otimes\sigma^{j}_{\rm 2}\big)e^{-i(H^{1,2}_{\rm XX})t}\big(\sigma^{l}_{\rm 1}\otimes \sigma^{k}_{\rm 2}\big)e^{i(H^{1,2}_{\rm XX})t}\right] \,.
\end{equation}
The Hamiltonian parameters enter in the unitary evolution in the following combinations:
\begin{equation}
\label{eq:Heigen}
    \begin{split}
&h_{12}=\frac{h_{1}+h_{2}}{2}\,,\,\,\, \quad \omega_{12}={\rm sgn}(\Delta_{12})\sqrt{\Delta_{12}^{2}+J_{12}^{2}}\\
&\Delta_{12}=h_{1}-h_{2}\,,\,\,\, \quad \tan\phi_{12}=\frac{J_{12}}{\Delta_{12}}\,.
        \end{split}
\end{equation}
Further details can be found in Appendix D. Then the evolution of the first spin in the first pair, for example, is given by
\begin{equation}
    \rho_{1}(t) ={\rm tr}_{2}\left[ e^{-i (H^{1,2}_{\rm XX}) t}\rho_{1}(0)\otimes \rho_{2}(0)  e^{i (H^{1,2}_{\rm XX}) t} \right] \equiv \Lambda_{1}(t,0) \rho_{\rm 1}(0) \,.
\end{equation}
Using the Bloch parameterization for the second spin (which provides the initial environment state for the first spin) as $\rho_2=\frac{1}{2}(1+x_2X_2+y_2Y_2+z_2Z_2)$, and defining partial components
\begin{equation}
\Lambda_{1}^{ijk}=\mathcal{U}^{i0}_{jk} \,,
\end{equation} 
the dynamical map in the Pauli basis is
\begin{equation}
\label{eq:XXLambda1}
\begin{split}
    &\Lambda_{1}(t,0)=\\&\begin{bmatrix} 1&0&0&0 \\ 0&\Lambda_{1}^{xx0} &\Lambda_{1}^{xy0}& \big(x_{2}\Lambda_{1}^{xzx}{\rm +}y_{2}\Lambda_{1}^{xzy}\big)\\ 0&\Lambda_{1}^{yx0} &\Lambda_{1}^{yy0}& \big(x_{2}\Lambda_{1}^{yzx}{\rm +}y_{2}\Lambda_{1}^{yzy}\big)\\ z_{2}\Lambda_{1}^{z0z} &\big(x_{2}\Lambda_{1}^{zxx}{\rm +}y_{2}\Lambda_{1}^{zxy}\big)&\big(x_{2}\Lambda_{1}^{zyx}{\rm +}y_{2}\Lambda_{1}^{zyy}\big) & \Lambda_{1}^{zz0} 
  \end{bmatrix}
  \end{split}
\end{equation}
and is non-unital so long as $z_{\rm 2}(0)\neq0$. The dynamical map $\Lambda_{2}(t,0)$ for the second spin can be obtained directly from Eq.(\ref{eq:XXLambda1}). From the Hamiltonian it is easy to see that we need only exchange the values of $h_i$, so $\Lambda_{2}(t,0)$ is directly obtained by exchanging the initial states and sending $\Delta_{12} \rightarrow -\Delta_{12}$. In terms of map components, this is accomplished by
\begin{equation}
    \Lambda_{2}^{ijk}(t,\Delta_{12})=\mathcal{U}^{0i}_{kj}(t,\Delta_{12})=\mathcal{U}^{i0}_{jk}(t,{\rm -}\Delta_{12})=\Lambda_{1}^{ijk}(t,{\rm -}\Delta_{12})\,.
\end{equation}
The dynamical maps associated to the second qubit pair are obtained through the replacement $12 \rightarrow 34$ in the initial states, $h_i$ and $J_{i,i+1}$. In this way we obtain an ensemble of non-unital 1-qubit dynamical maps and unitary 2-qubit dynamical maps. 

Given the collection of disjoint qubit pairs, and the associated 2-qubit unitary channels, it is possible to construct their ensemble averages. Both averages will be non-unitary, since convex combinations of unitary channels may be non-unitary. This provides a useful means to efficiently simulate non-unitary evolution \cite{Cleve:2016dgx}. If each pair is assumed to be in the same uncorrelated 2-qubit state, the following non-unitary ensemble averaged 2-qubit dynamical map may be defined 
\begin{equation}
    \overline{\mathcal{U}}(t)\left[\rho_{1}(0)\otimes\rho_{2}(0)\right]=\frac{1}{N_{\rm CL}}\sum_{I=1}^{N_{\rm CL}}\left(e^{-i \left(H^{I}_{\rm XX}\right) t}\rho_{1}(0)\otimes \rho_{2}(0) e^{i \left(H^{I}_{\rm XX}\right) t} \right) \,.
\end{equation}
At this point we have a physical representation of the sort of noisy, or disordered, Hamiltonians that have the same structure but with couplings drawn from some distribution. Now conceptually we may view the the previous ensemble as an instantiation of some noise distribution over Hamiltonian parameters. Therefore in the limit of large $N_{\rm CL}$ the network average is equivalent to 
\begin{equation}
    \overline{\mathcal{U}}(t)\left[\rho_{1}(0)\otimes\rho_{2}(0)\right]=\langle e^{-i H_{\rm XX} t}\rho_{1}(0)\otimes \rho_{2}(0) e^{i H_{\rm XX}t}\rangle_{\rm disorder}
\end{equation}
where we have dropped the superscripts on $H_{\rm XX}$ to emphasis the conceptual difference with the network of clusters. 

 Now instead of considering a disorder distribution directly over the parameters in the Hamiltonian, we consider the distributions directly over its eigenvalues and eigenvectors. That is, we take the fundamental parameters to be those on the left hand side of Eq.(\ref{eq:Heigen}), e.g. $h$ (from $h_{12})$, $\omega$, and $\phi$, rather than $(h_{1},h_{2},J_{\perp})$. This approach allows for a closer connection to the dynamical map components. For example, the following component
\begin{equation}
\label{eq:xzyconstraint}
    \Lambda^{xzy}_{1}\propto \sin\phi \,,
\end{equation}
which is compactly expressed in terms of the mixing angle $\phi$ from $H_{\rm XX}$ eigendecomposition. Notice that such a component vanishes if the noise distribution is an even function of $\phi$, where $\tan\phi=\frac{J}{\Delta}$. The other dynamical map components that break phase-covariance also vanish under the same assumption. We understand this as the phase-covariant constraints in Eq.(\ref{eq:PCcond}) being satisfied on average. 

We can apply a similar procedure to other non-phase-covariant maps. In particular, for more general non-unital quantum channels with non-zero shift in the $x$ and $y$ direction ($\tau_1\neq0$ and $\tau_2 \neq0$), the invariant state can have coherences and therefore cannot be represented as a Gibbs state in the computational basis. However, ensembles of such general non-unital maps can be generated such that the shift in the $x$ and $y$ directions have mean zero, resulting in a Gibbs-like state for the ensemble-averaged map.

\subsection{Phase-covariance via disorder averaging}
We begin by considering independent Gaussian distributions for $h$ and $\omega$,
\begin{equation}
p_{G}(h)= \frac{1}{\sqrt{2\pi\sigma_{h}^{2}}}e^{-\frac{(h-B)^{2}}{2\sigma_{h}^{2}}}\,\,\,,\,\,\,\,\,\,p_{G}(\omega)= \frac{1}{\sqrt{2\pi\sigma_{\omega}^{2}}}e^{-\frac{(\omega-\Omega)^{2}}{2\sigma_{\omega}^{2}}}\,.  
\end{equation}
which have average values $B$ and $\Omega$ and variances $\sigma_{h}$ and $\sigma_{\omega}$ respectively. 
We use Gaussian distributions as the simplest first choice, where only two parameters must be specified. The $\phi$ distribution requires a bit more care. In order to ensure that the average single-spin dynamical map is phase-covariant,  any distribution must enforce $\langle \phi \rangle=0$ (as explained below Eq.(\ref{eq:xzyconstraint})). With this constraint, the Gaussian distribution has only a single free parameter, namely its variance $\sigma_{\phi}$:
\begin{equation}  
   p_{\rm G}(\phi)= \frac{1}{\sqrt{2\pi\sigma_{\phi}^{2}}}e^{-\frac{\phi^{2}}{2\sigma_{\phi}^{2}}} \,\,\,,\,\,\,\,\,\,
\end{equation}
However, notice that the average value of $\tau_{3}$, assuming $\phi$ obeys a Gaussian distribution, is
\begin{equation}
    \begin{split}
\overline{\tau}_{3}\Big|_{ p_{G}(\phi)}=z_{2}\overline{\Lambda}_{1}^{z0z}=\frac{z_{2}}{4}\big(1-e^{-2\sigma^{2}_{\phi}}\big)\big(1-\cos{(2\Omega t)}e^{-2\sigma_{\omega}^{2}t^{2}}\big)\,
\end{split}
\end{equation}
which saturates to a maximum value $\frac{z_{2}}{4}$ if $\sigma_{\phi}=0$. Therefore the choice of a Gaussian distribution fixes a maximal size of the non-unitality to be $\frac{1}{4}$. But as one can determine from our exact ensembles, there is no reason to put such a limitation on the size of the non-unitality. We may choose a different distribution that allows larger non-unitality, for example
\begin{equation}
p_{\rm T}(\phi)= \frac{\tanh[a_{\phi}\phi]^2}{\left(\pi-\frac{2}{a_{\phi}}\tanh\left[\frac{a_{\phi}\pi}{2}\right]\right)}\,,\;\;\,
\end{equation}
where the range is restricted to be $-\pi/2\leq\phi\leq\pi/2$. The parameter $a_{\phi}$ controls how quickly the distribution saturates to its maximum value. For small values of $a_{\phi}$ the distribution reaches its maximum value only near the ends of the interval $[-\frac{\pi}{2},\frac{\pi}{2}]$, where as for large values the distribution saturates much more closely to $\phi=0$. Using the truncated hyperbolic tangent distribution ($p_{T}(\phi)$) one finds that 
\begin{align}
    \lim_{a_{\phi}\rightarrow 0}\langle \sin^2[\phi] \rangle \Big|_{p_{T}(\phi)} = \frac{6+\pi^2}{2\pi^2}\approx 0.8 \,.
\end{align}
is the maximum allowed value for the average over the $\phi$-dependent part of $\tau_{3}$.  This leads to a maximum possible saturation value of $\overline{\tau}_{3}$ of $\frac{2}{5}z_{2}$. Therefore by adjusting the shape of the $\phi$-distribution a larger set of phase-covariant maps may be reached after averaging. In addition, by altering the choice of the $h$ and $\omega$ distributions, this value can be made larger, although we do not pursue such examples here. 

%%%%%%%%%%%%%%%%%%%%%%%%%%%%%%%%%%%%%%%%%%%%%%%%%%%%%%%%%%%%%%%
%%%%%%%%%%%%%%%%%%%%%%%%%%%%%%%%%%%%%%%%%%%%%%%%%%%%%%%%%%%%%%%
\section{Phase-covariant dynamics from channel ensembles}
\label{sec:Construction}
\setcounter{equation}{0}
%%%%%%%%%%%%%%%%%%%%%%%%%%%%%%%%%%%%%%%%%%%%%%%%%%%%%%%%%%%%%%%
%%%%%%%%%%%%%%%%%%%%%%%%%%%%%%%%%%%%%%%%%%%%%%%%%%%%%%%%%%%%%%%
In the final section of this work we construct measures over phase-covariant channels that capture key features of the long-time averaged channels derived from XXZ networks. However, the time-evolving ensemble can be considered on its own, characterized by its thermodynamic properties without reference to a particular Hamiltonian. From the Gaussian examples considered in Section 4, we are motivated to consider a phase-covariant measure consisting of truncated Gaussian distributions over the non-unitary parameters. We use the physical constraints from Section 2 to fix the mean ($\mu$) and scale ($\sigma$) parameters. 

%%%%%%%%%%%%%%%%%%%%%%%%%%%%%%%%%%%%%%%%%%%%%%
\subsection{Phase-covariant measures}
\label{sec:PCensembles}
%%%%%%%%%%%%%%%%%%%%%%%%%%%%%%%%%%%%%%%%%%%%%%
The key point in defining a measure over channels is determining the domain of integration. This depends on the constraints imposed by complete positivity, which are especially simple in the case of phase-covariant qubit channels. Then, a phase-covariant measure may be defined by supplying a joint probability distribution $p_{N}(\lambda_{1},\tau_{3},\lambda_{3};t)$ which satisfies the normalization condition
\begin{equation}
\begin{split}
 1=&\int_{\Gamma_{\rm PC}} p_{N}(\lambda_{1},\tau_{3},\lambda_{3};t) d\lambda_{1} d\tau_{3} d\lambda_{3} \\ =&\int_{0}^{1}\int_{-(1-\lambda_{3})}^{1-\lambda_{3}}\int_{-\frac{1}{2}\sqrt{(1+\lambda_{3})^{2}-\tau_{3}^{2}}}^{\frac{1}{2}\sqrt{(1+\lambda_{3})^{2}-\tau_{3}^{2}}}p_{N}(\lambda_{1},\tau_{3},\lambda_{3};t)d\lambda_{1} d\tau_{3} d\lambda_{3} \\+& \int_{-1}^{0}\int_{-(1+\lambda_{3})}^{1+\lambda_{3}}\int_{-\frac{1}{2}\sqrt{(1+\lambda_{3})^{2}-\tau_{3}^{2}}}^{\frac{1}{2}\sqrt{(1+\lambda_{3})^{2}-\tau_{3}^{2}}}p_{N}(\lambda_{1},\tau_{3},\lambda_{3};t)d\lambda_{1} d\tau_{3} d\lambda_{3}\,,
\end{split}
\end{equation}
where the order of integration follows from the form of the constraints in Eq.(\ref{eq:CP2}). For more general quantum systems it may not be possible to determine the boundary of integration, but restricting the reduced dynamics to be G-covariant potentially makes the problem tractable for larger quantum channel. 

The above measure allows us to define the average dynamical map,    
\begin{equation}
\langle\Lambda_{N}\rangle(t)=\int_{\Gamma_{\rm PC }}p_{N}(\lambda_{1},\tau_{3},\lambda_{3};t) \Lambda(\lambda_{1},\tau_{3},\lambda_{3})d\lambda_{1} d\tau_{3} d\lambda_{3} \,. 
\end{equation}
 We choose to introduce time-dependence into the probability distribution, instead of directly into the dynamical map, as the time-dependence of the dynamical map becomes quite complicated for large clusters. 

This form of the map ensemble can be usefully compared to prior explorations of random open system dynamics, and to compare the phase-covariant class to the larger class of channels. 
  
  We compare the distribution of eigenvalues for a non-unital phase-covariant ensemble to those of the more general class of dynamical maps, treated for example in \cite{Bruzda:2009}. To do so, it is helpful to diagonalize the phase-covariant channel and define
 \begin{align}
D_{\Lambda_{\rm PC}}=
     \begin{bmatrix} 
     1 & 0 & 0 & 0 \\ 
     0 & \lambda_{1} e^{i \theta} & 0 & 0 \\ 
     0 & 0 & \lambda_{1} e^{-i \theta} & 0 \\
     0 & 0 & 0 & \lambda_{3} \\
     \end{bmatrix}\,.
\end{align}
Note that eigenvalues of any phase-covariant channel do not depend on $\tau_{3}$, the stationary state does as we have
\begin{equation}
    \rho_{\star}=\frac{1}{2}(\mathbb{1}+\frac{\tau_{3}}{1-\lambda_{3}}Z)\,.
\end{equation}
The phase-covariant channel parameters are uniformly sampled over the completely-positive regions and the associated eigenvalues plotted on the left panels of Figures 6 and 7 where $\mu_{\pm}=\lambda_{1} e^{\pm i \theta}$. The complexity of the eigenvalues follows from the presence of the rotation, which describes the unitary part of the dynamics. The complex phases depend only on the rotation angle, and so when it is not included in the definition of phase-covariant dynamics, the eigenvalues will be real (see, e.g. \cite{Siudzinska:2023a}). As this is not the case for more general channels, it is useful to explicitly include the rotation angle here.

To make a comparison with the more general class of dynamical maps in \cite{Bruzda:2009}, we consider a minimal breaking of the phase-covariance through the introduction of a single new parameter, $\lambda_2$, without allowing any new non-zero components in the map:
\begin{equation}
        \Lambda_{\rm PC} \xrightarrow[\text{PC}]{\text{Break}}
         \begin{bmatrix} 
     1 & 0 & 0 & 0 \\ 
     0 & \lambda_{1} \cos{\theta} & -\lambda_{2} \sin{\theta} & 0 \\ 
     0 & \lambda_{1} \sin{\theta} & \lambda_{2} \cos{\theta} & 0 \\
     \tau_{3} & 0 & 0 & \lambda_{3} \\
     \end{bmatrix}\,.
\end{equation}
This is of course not the only way to break phase-covariance. For the spin systems considered in previous sections, considering initial states that do not satisfy the phase-covariant restriction (second equality of Eq.(\ref{eq:PCcond})) but keeping the Hamiltonian fixed maintains the simple structure of the central components of the map but introduces new non-zero terms that depend on the initial state. So, the simple breaking chosen above can be tied instead to breaking the symmetry of the interacting Hamiltonian rather than the choice of initial state.

The eigenvalues of the non-phase-covariant map above are
\begin{equation}
    \mu_{\pm} = \frac{(\lambda_1+\lambda_2)\cos{\theta}\pm\sqrt{(\lambda_1+\lambda_2)^2\cos^2{\theta-4\lambda_1\lambda_2}}}{2}.
\end{equation}
The dynamical map parameters are uniformly sampled (again restricting to complete positivity) and the eigenvalues of the non-phase-covariant ensemble are plotted on the right panels of Figures 6 and 7. Notice for these maps phase of the complex eigenvalues depends not only on $\theta$ but also $\lambda_{1}$ and $\lambda_{2}$.

Even with this simple breaking of phase covariance, additional structures may seen in Figure 6 (right), several of which match those in Figure 1 of the slightly more general treatment of \cite{Bruzda:2009}. For example, there is agreement in the empty parts of the plane, and the apparent `repulsion' of the points away from the real axis. Note also that a significant number of $\mu_{\pm}$ eigenvalues become real when phase covariance is broken. As we have used the same number of total points in each panel, the density of points off the real axis on the right panel is lower than on the left, and we have plotted the real $\lambda_3$ eigenvalues in the foreground.
 \begin{figure}[H]
\centering
\includegraphics[width=1\linewidth]{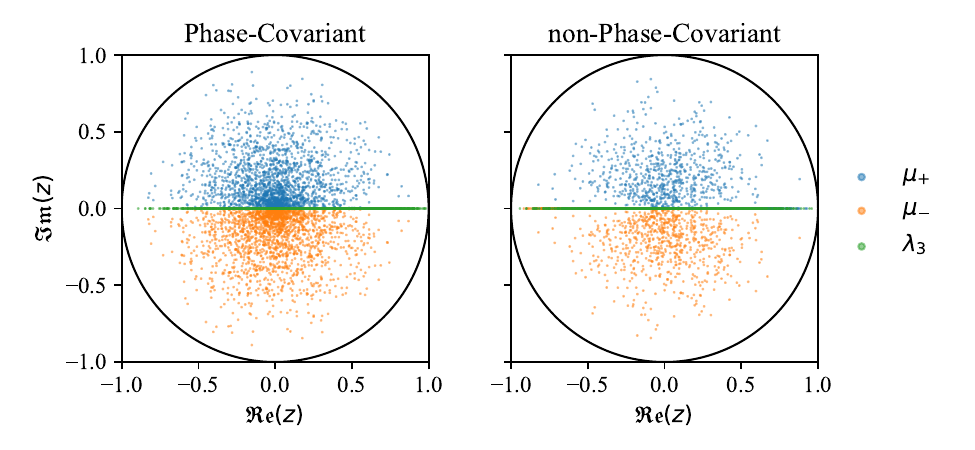}
\caption{The distribution of eigenvalues, $\mu_{\pm}=\lambda_{1} e^{\pm i \theta}$, $\lambda_{3}$, for a non-unital phase-covariant ensemble (left) and non-unital non-phase-covariant ensemble (right) in the complex plane.}
\label{fig:eigendis}
\end{figure}

 \begin{figure}[H]
\centering
\includegraphics[width=1\linewidth]{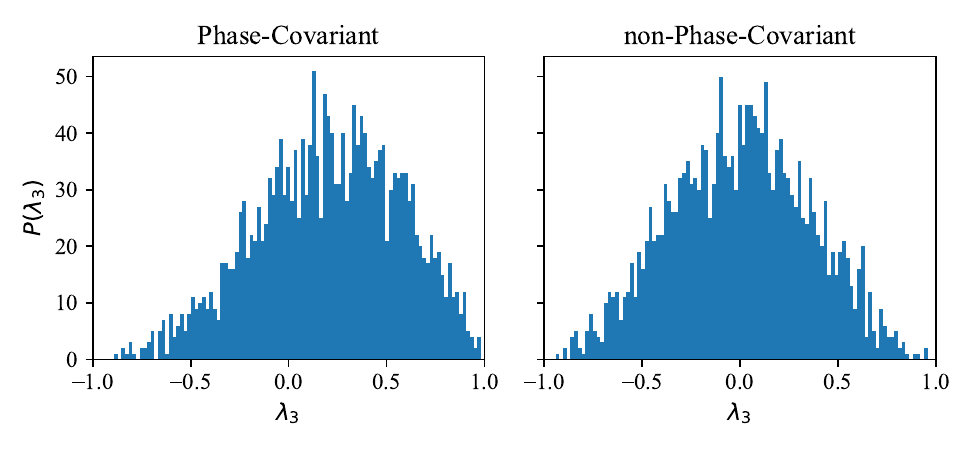}
\caption{Plotted are histograms of the distribution of the eigenvalue, $\lambda_{3}$, for a non-unital phase-covariant ensemble (left) and non-unital non-phase-covariant ensemble (right). \label{fig:lambda3dis}}
\end{figure}
Finally, we end by defining the uniform distribution over phase-covariant channels by computing the volume of phase-covariant channels finding
\begin{equation} 
\label{eq:bias}
vol\big(\Gamma_{\rm PC}\big)= vol\big(\Gamma^{+}_{\rm PC}\big)+vol\big(\Gamma^{-}_{\rm PC}\big)=\left(\frac{16}{9}-\frac{\pi}{6}\right)+\frac{\pi}{6}=\frac{16}{9} \,.
\end{equation}
A uniform distribution is then defined with probability density $\frac{9}{16}$.
Here we have explicitly  separated the integral into two regions based on the sign of $\lambda_{3}$. The positive region has a larger volume than that of the negative region, which is confirmed in Figure 7 (left). Therefore when sampling the phase-covariant channels using a uniform distribution, we expect there to be a bias towards positive $\lambda_{3}$.

In \cite{Siudzinska:2023a} the volume of phase-covariant channels was obtained using the volume form
\begin{equation}
    dV=\frac{\sqrt{2}}{8}d\lambda_{1}d\lambda_{3}d
    \tau_{3}\,
\end{equation}
obtaining $\frac{2\sqrt{2}}{9}$ as the volume of phase-covariant channels. The volume we have obtained in Eq.(\ref{eq:bias}) is consistent with this result, but simply rescaled by a factor of $\frac{8}{\sqrt{2}}$. Obviously, such a re-scaling has no effect on the relative volumes of $\Gamma^{+}_{\rm PC}$ and $\Gamma^{-}_{\rm PC}$. 
\subsection{Example distribution: Truncated-Gaussians}
 There are of course many other non-uniform distributions to use to define a phase-covariant measure, and making an appropriate choice allows one to more easily capture the behavior of the long time average channels determined in Section 2. Although a time-dependent uniform distribution could be used, this is a more difficult to define from the data gathered from the average and late-time fluctuations. Ostensibly, the results of Section 4 indicate that using truncated Gaussian distributions could be a fruitful approach. We consider phase-covariant measures defined in this way for the remainder of this section.

A truncated Gaussian distribution over a single random variable ($X$) is given by 
\begin{equation}
    p_{\rm TG}(X;\vec{\eta})=\frac{e^{-\frac{(X-\mu)^{2}}{2\sigma^{2}}}}{\mathcal{N}\left(\vec{\eta}\right)}
\end{equation}
where $\vec{\eta}$ stands in for the parameters of the distribution and $\mathcal{N}$ is the normalization factor
\begin{equation}
\mathcal{N}\left(\vec{\eta}\right)=\int_{a}^{b}e^{-\frac{(X-\mu)^{2}}{2\sigma^{2}}}dX \,.
\end{equation}
There are four total parameters associated to a two-sided truncated Gaussian, the mean parameter ($\mu$) , the scale parameter $(\sigma)$, and the end points of the interval that define the truncation. For our purposes the end points of each distribution are set by the complete-positivity constraints. For example the truncated Gaussian associated with $\lambda_{3}$ may take values between -1 and 1, while the other distributions have intervals depending on the values of $\lambda_{3}$. 

The remaining parameters we set through the use of the physical constraints found in the previous section. We define the mean of each parameter by
\begin{equation}
    \begin{split}
        &\mu_{1}(N)=\overline{\lambda}^{\infty}_{1}(N)=0\\ 
        &\mu_{\lambda}(N)=\overline{\lambda}^{\infty}_{3}(N) \\ 
        &\mu_{\tau}(N)=\overline{\tau}^{\infty}_{3}(N)\,,
    \end{split}
\end{equation}
where as an approximation one may take $\mu_{\lambda}\approx a_{0}$ and $\mu_{\tau}\approx (1-a_{0})\overline{z}$, where $a_{0}$ is the first term in the expansion of $\overline{\lambda}_{3}(N)$ in the polynomials of the initial state. We set the size of typical fluctuations using our exact computations set the size of typical fluctuations, although there is not unique way to do this. For a Gaussian distribution, a natural choice is to set the variance $\sigma$ to be proportional to the envelope function determined above (see Figure 3 or Figure 4). For example, setting the envelope to correspond, approximately, to 3-sigma fluctuations, the scale parameter (assumed real) takes the form
\begin{equation}
\label{eq:scaleparameter}
\sigma=\frac{1}{3}\frac{C}{N}\frac{ t_{\rm ref}}{t}\,,
\end{equation}
where the constant $C$ is $O(1)$. A specific value of $C$ can be obtained for a particular choice of initial state of known Hamiltonian and size. But, more generally one should think of a specified coefficient of $\frac{ t_{\rm ref}}{t}$ in $\sigma$ as folding in some degenerate information about the initial state, the size of the system, and a choice of precision in the statistics of the underlying system. Notice that since we have used truncated Gaussian distributions, these parameters do not correspond to the exact mean and variance, and in general are larger. 

In order to construct instances of these distributions and how they change over time we use the following procedure. First, we choose a sampling time, set by the time-scale $t_{\rm ref}=t_{J}=\frac{2\pi}{J_{\perp}}$. We then sample the distribution $p_{\rm TG}\left(\lambda_{3},\vec{\eta}_{3}\right)$ at each time step, where the scale parameter is changed according to Eq.(\ref{eq:scaleparameter}). In this way we obtain an ordered list of values that $\lambda_{3}$ during each time-step. From these values of $\lambda_{3}$, we may construct the distribution $p_{\rm TG}(\tau_{3},\vec{\eta}_{\tau})$, as recall the end-point parameters for a given time-step depend explicitly on obtained values of $\lambda_{3}$. The procedure then carries on in the same manner as for $\lambda_{3}$, a successive list of values for $\tau_{3}$ are generated for each time-step. Finally, using the outcomes of the previous steps a list of values for $\lambda_{1}$ may be generated from the distribution $p_{\rm TG }(\lambda_{1},\vec{\eta}_{1})$. Figure 7  illustrates the results of this procedure. One could define a smaller sampling time, as it is clear that the sampling procedure does not capture the smaller scale fluctuations. However, these fluctuations do not carry additional physical information about the convergence toward the steady channel, for single-qubit dynamics, at least. 

Care must be taken into the above procedure, as the endpoints for the distribution over $\tau_{3}$ depends explicitly on the sign of $\lambda_{3}$. One way around the potential issues is to simply take the $\tau_{3}$-interval to be set by the sign of the late-time limit of $\lambda_{3}$. This guarantees that at late-enough times, no positivity breaking occurs. Of course one may just consider the sign of $\lambda_{3}$ in the above procedure and take care to use the correct interval for $\tau_{3}$ at a given time-step.

\begin{figure}[H]
\centering
\begin{subfigure}{.45\linewidth}
  \centering
\includegraphics[width=0.95\linewidth]{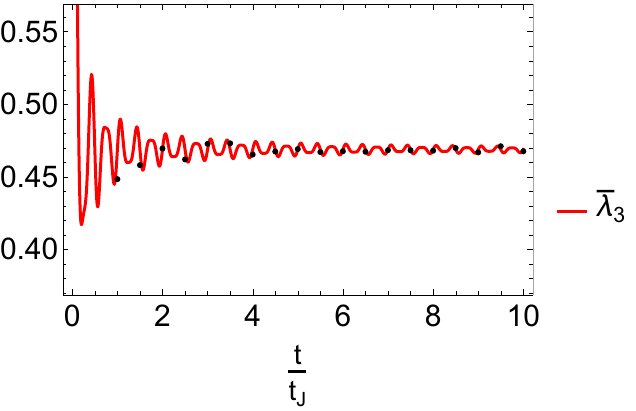} 
\caption{Homogeneous I.C.}
\end{subfigure} 
\begin{subfigure}{.45\linewidth}
  \centering
\includegraphics[width=0.95\linewidth]{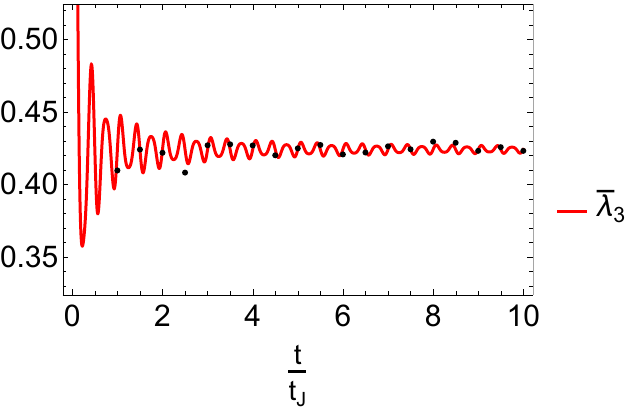} 
 \caption{N\'{e}el I.C.}
\end{subfigure}
\caption{The time and network averaged $\lambda_{3}(t)$ are plotted for the completely-connected XXZ-cluster with 5 spins. The left plot is generated assuming an homogeneous initial condition $z_{i}$=0.1, and the right plot is generated using a N\'{e}el state. Included within these plots are instantiations of the time-dependent truncated Gaussians plotted, represented as the black dots within each plot. We see from these plots that the distributional approach captures the dynamics well especially at late times $\frac{t}{t_{J}}\rightarrow \infty$. We have considered these two initial states to demonstrate that taking $C$ to be initial state independent as in Eq.(\ref{eq:scaleparameter}) still results in a distribution that well approximates the averaged dynamics.}
\label{fig:TGL3CC}
\end{figure}

%%%%%%%%%%%%%%%%%%%%%%%%%%%%%%%%%%%%%%
%%%%%%%%%%%%%%%%%%%%%%%%%%%%%%%%%%%%%%
\section{Conclusions}
\label{sec:conclude}
In this work we have defined phase-covariant ensembles in two different ways. The first approach was to construct ensembles directly from a time-independent Hamiltonian, and consider the individual reduced dynamics for all possible subsystems of the smallest size. We constructed examples where each dynamical map in the ensemble was phase-covariant, and examples where only the network averaged reduced dynamics was phase-covariant. From these examples, we are able to conjecture the form of the map parameters in terms of the initial state and connectivity, as shown in Eq.(\ref{eq:ccQ4}) and Eq.(\ref{eq:ccQ5}) for completely connected maps and Eq.(\ref{eq:ring4}) and Eq.(\ref{eq:ring5}) for one-dimensional rings with only nearest neighbor interactions. While we are unable to conjecture the general form of the long-time average for the ring connectivity, we do see that in general only the translation symmetry should be imposed. 

Grounded in those examples, we proposed a method to generate ensembles of arbitrary size via distributions of parameters for dynamical maps. The distributions were constructed so that the ensemble of maps would be constrained such that the individual open systems could be considered a partition of a larger, closed system. More specifically, the distributions were built to enforce some average dynamics carrying properties of the Hamiltonian and initial state, with fluctuations typical of the variation of different environmental initial states seen by each subsystem.

Ideally, one would like to understand how all known properties of the closed system are evident in the ensemble of open-system dynamics of subsystems. However, in the examples we studied the phase-covariant dynamics of 1-qubit subsystems do not seem to differ based on the integrable vs. non-integrable nature of the spin-chain Hamiltonian considered. In principle, such a characterization should be crucially important in determining the nature of a random distribution over channels, similar to how it constrains the form of random unitary dynamics in various large systems. It will be interesting to expand this study to both larger subsystems and more explicit breaking of phase-covariance.

There are a number of implications of these results that are specific to the use of phase-covariant dynamics. For example, the resulting time-independent, phase-covariant channels generate a notion of thermalization associated with each Hamiltonian and initial state, since under many repeated applications of the steady channel, every single-spin state will be driven toward the Gibbs state associated with $\beta_* = \log \Big[ \frac{2}{1-\overline{a}^{\infty}_*}\Big]$, where $\overline{a}^{\infty}_* = \frac{\overline{\tau}^{\infty}_3}{1-\overline{\lambda}^{\infty}_3}$. The repeated application of such maps has also found a role in the simulation of open quantum systems \cite{Pocrnic:2023}, although so far restricted to systems described by stationary Lindblads. Our results may open up a similar route to efficient simulation of a broader class of open systems. This is especially interesting as quantum simulators may have an exponential advantage for simulating open-system evolution \cite{Kashyap:2024wgf}. The procedure in Section 5 could be used in this direction.

%%%%%%%%%%%%%%%%%%%%%%%%%%%%%%%%%%%%%%
%%%%%%%%%%%%%%%%%%%%%%%%%%%%%%%%%%%%%%

%%%%%%%%%%%%%%%%%%%%%%%%%%%%%%%%%%%%%%
%%%%%%%%%%%%%%%%%%%%%%%%%%%%%%%%%%%%%%
\section{Acknowledgements}
This work was supported by NSF PHY-2310662.

%%%%%%%%%%%%%%%%%%%%%%%%%%%%%%%%%%%%%%
%%%%%%%%%%%%%%%%%%%%%%%%%%%%%%%%%%%%%%

\appendix

\section*{Appendix A}
\label{app:A}
\def\theequation{A.\arabic{equation}}
\setcounter{equation}{0}

In this appendix we lay out the symmetries that are useful in computing exact dynamical maps. Throughout the paper, we use the Pauli matrices $X,Y,Z$, defined by
\begin{equation}
\label{eq:Paulis}
    X=\begin{bmatrix}0&1\\1&0\end{bmatrix},\;\;\;Y=\begin{bmatrix}0&-i\\i&0\end{bmatrix},\;\;\;Z=\begin{bmatrix}1&0\\0&-1\end{bmatrix}\,.
\end{equation}

The first of these symmetries is the charge, or equivalently the excitation number, operator
\begin{equation}
Q_{N}=\sum_{i}Z_{i} \,.
\end{equation} 
Throughout the article the excitation number is labeled as $q$, but note this is not the eigenvalue of $Q_{N}$. Each block with a given value of $q$ has dimension $m=$$N\choose q$, simply the number of ways to pick $q$ qubits to be excited.

  Hamiltonians satisfying $[H,Q_{N}]=0$ will be block diagonal in any excitation eigenbasis. Since $Q_{N}$ is already diagonal in the computational basis, block diagonalizing the Hamiltonian at most requires a permutation matrix that reorders computational basis states by excitation number. For this article we chose $\mathcal{P}_{N}$ so that the computational states within the same $q$-blocks are ordered based on their binary representation. For example when $N=3$, in the $q=1$ block the computational state $|001\rangle$ comes before $|100\rangle$. The next symmetry available to the models we consider is the translation operator $(T_{N})$. The translation action on computational basis states is
\begin{equation}
T_{N}|a_{1}a_{2}...a_{N}\rangle=|a_{N}a_{1}...a_{N-1}\rangle
\end{equation}
and the adjoint action on Pauli matrices is
\begin{equation}
T_{N}\sigma^{\alpha}_{i}T_{N}^{\dag}=\sigma^{\alpha}_{i+1} \,.
\end{equation}
where a tensor product of the Pauli with the identity operator on all other spins is implied. Since $[Q_{N},T_{N}]=0$ the translation operator is block diagonal in any excitation-ordered computational basis. Further noting that $T_{N}^{N}=\mathbb{1}$, it is not hard to see that its eigen-spectrum consists entirely of $N^{\rm th}$ roots of unity.  Combining these results we may label the eigenstates of  $T_{N}$ as
\begin{equation}
\begin{split}
&T_{N}|\mathcal{F}^{a}_{q};k\rangle=e^{i\frac{2\pi a}{N}}|\mathcal{F}^{a}_{q};k\rangle\\ 
&Q_{N}|\mathcal{F}^{a}_{q};k\rangle=\big(2q-N\big)|\mathcal{F}^{a}_{q};k\rangle
\end{split}
\end{equation}
where $q$ is the excitation number, and $k$ is a label that removes any remaining degeneracy of the eigenstates. In a given $q$-block, Fourier modes corresponding to subgroups of the cyclic group $\mathbb{Z}_{N}$ may appear, which happens when $N$ is a composite integer. In the remainder of this appendix we give constructions of the Fourier modes for $N=3$ and $N=4$. 
\subsection*{3-qubits}
\begin{equation}
    \begin{split}
&|\mathcal{F}^{0}_{0};0\rangle=|111\rangle \quad |\mathcal{F}^{0}_{3};0\rangle=|000\rangle
\end{split}
\end{equation}
\begin{equation}
\begin{split}
&|\mathcal{F}^{0}_{1};0\rangle=\frac{1}{\sqrt{3}}\Big(|011\rangle+|101\rangle+|110\rangle \Big)\\
&|\mathcal{F}^{1}_{1};0\rangle=\frac{1}{\sqrt{3}}\Big(|011\rangle+e^{\frac{2\pi i}{3}}|101\rangle+e^{\frac{4\pi i}{3}}|110\rangle\Big) \\
&|\mathcal{F}^{2}_{1};0\rangle=\frac{1}{\sqrt{3}}\Big(|011\rangle+e^{\frac{4\pi i}{3}}|101\rangle+ e^{\frac{2\pi i}{3}}|110\rangle\Big)
\end{split}
\end{equation}
\begin{equation}
\begin{split}
&|\mathcal{F}^{0}_{2};0\rangle=\frac{1}{\sqrt{3}}\Big(|001\rangle+|010\rangle+|100\rangle \Big)\\
&|\mathcal{F}^{1}_{2};0\rangle=\frac{1}{\sqrt{3}}\Big(|001\rangle+e^{\frac{2\pi i}{3}}|010\rangle+e^{\frac{4\pi i}{3}}|100\rangle\Big) \\
&|\mathcal{F}^{2}_{2};0\rangle=\frac{1}{\sqrt{3}}\Big(|001\rangle+e^{\frac{4\pi i}{3}}|010\rangle+ e^{\frac{2\pi i}{3}}|100\rangle\Big)
\\ \\
\end{split}
\end{equation}
\subsection*{4-qubits}
\begin{equation}
    \begin{split}
&|\mathcal{F}^{0}_{0};0\rangle=|1111\rangle \quad |\mathcal{F}^{0}_{4};0\rangle =|0000\rangle
\end{split}
\end{equation}
\begin{equation}
\begin{split}
&|\mathcal{F}^{0}_{1};0\rangle=\frac{1}{\sqrt{4}}\Big(|0111\rangle+|1011\rangle+|1101\rangle+|1110\rangle \Big)\\
&|\mathcal{F}^{1}_{1};0\rangle=\frac{1}{\sqrt{4}}\Big(|0111\rangle+i|1011\rangle-|1101\rangle-i|1110\rangle\Big) \\
&|\mathcal{F}^{2}_{1};0\rangle=\frac{1}{\sqrt{4}}\Big(|0111\rangle-|1011\rangle+ |1101\rangle-|1110\rangle\Big)\\
&|\mathcal{F}^{3}_{1};0\rangle=\frac{1}{\sqrt{4}}\Big(|0111\rangle-i|1011\rangle-|1101\rangle+i|1110\rangle\Big)
\end{split}
\end{equation}
\begin{equation}
\begin{split}
&|\mathcal{F}^{0}_{3};0\rangle=\frac{1}{\sqrt{4}}\Big(|0001\rangle+|0010\rangle+|0100\rangle+|1000\rangle \Big)\\
&|\mathcal{F}^{1}_{3};0\rangle=\frac{1}{\sqrt{4}}\Big(|0001\rangle+i|0010\rangle-|0100\rangle-i|1000\rangle\Big) \\
&|\mathcal{F}^{2}_{3};0\rangle=\frac{1}{\sqrt{4}}\Big(|0001\rangle-|0010\rangle+ |0100\rangle-|1000\rangle\Big)\\
&|\mathcal{F}^{3}_{3};0\rangle=\frac{1}{\sqrt{4}}\Big(|0001\rangle-i|0010\rangle-|0100\rangle+i|1000\rangle\Big)\\\\
\end{split}
\end{equation}
\begin{equation}
\label{eq:4Qextradegenerate}
\begin{split}
&|\mathcal{F}^{0}_{2};0\rangle =\frac{1}{\sqrt{4}}\Big(|0011\rangle+|0110\rangle+|1001\rangle+|1100\rangle \Big) \\
&|\mathcal{F}^{1}_{2};0\rangle= \frac{1}{\sqrt{4}}\Big(|0011\rangle+i|0110\rangle-|1001\rangle-i|1100\rangle \Big)\\
&|\mathcal{F}^{2}_{2};0\rangle=\frac{1}{\sqrt{4}}\Big(|0011\rangle-|0110\rangle+|1001\rangle-|1100\rangle \Big) \\
&|\mathcal{F}^{3}_{2};0\rangle= \frac{1}{\sqrt{4}}\Big(|0011\rangle-i|0110\rangle-|1001\rangle+i|1100\rangle \Big) \\
&|\mathcal{F}^{0}_{2};1\rangle =\frac{1}{\sqrt{2}}\big(|0101\rangle+|1010\rangle\big)\\ &|\mathcal{F}^{2}_{2};1\rangle=\frac{1}{\sqrt{2}}\big(|0101\rangle-|1010\rangle\big) 
\end{split}
\end{equation}
\section*{Appendix B}
\label{app:B}
\def\theequation{B.\arabic{equation}}
\setcounter{equation}{0}

In this Appendix we present the exact formulae for the dynamical maps considered in this work. We have the following form of phase-covariant maps
\begin{equation}
\label{eq:phasecovariant}
   \Lambda(t)=\begin{bmatrix}1&0&0&0 \\ 0&\Lambda_{xx}(t)&\Lambda_{xy}(t)&0 \\ 0&\Lambda_{yx}(t)&\Lambda_{yy}(t)&0 \\ \Lambda_{z0}(t)&0&0&\Lambda_{zz}(t)\end{bmatrix}
\end{equation}
where for the Hamiltonians considered, the various components all may be decomposed as 
\begin{equation}
\label{eq:nonunitary}
\begin{split}
    &\Lambda_{xx}(t)=\alpha(t) \cos{2ht}-\beta(t) \sin{2ht}=\lambda_{1}(t)\cos{(2ht+\phi(t))} \\
    &\Lambda_{yx}(t)=\beta(t)\cos{2ht}+\alpha(t) \sin{2ht}=\lambda_{1}(t)\sin{(2ht+\phi(t))} \\
    &\Lambda_{zz}(t)=\lambda_{3}(t) \\
    &\Lambda_{z0}(t)=\tau_{3}(t)\,, 
\end{split}
\end{equation}
with the remaining components determined by the relations $\Lambda_{yy}=\Lambda_{xx}$ and $\Lambda_{xy}=-\Lambda_{yx}$. 
To perform the minimum amount of computations, we only provide explicit formulas for $\tau_{3}$, $\lambda_{3}$, $\alpha$, and $\beta$ for each model. 

\subsection*{Completely-connected XXZ-network}
\label{sec:CCXXZ}

We present the exact dynamical map components for the completely-connected and homogeneous XXZ-network. We present all the non-unitary parameters defined in Eq.(\ref{eq:nonunitary}). 
\subsubsection*{4-qubits}
For 4-qubits the Hamiltonian has the block diagonal structure
\begin{equation}
\mathcal{P}_{4}H_{\rm XXZ}\mathcal{P}_{4}^{\dag}= H^{0}_{\rm XXZ}\oplus H^{1}_{\rm XXZ}\oplus H^{2}_{\rm XXZ}\oplus H^{3}_{\rm XXZ} \oplus H^{4}_{\rm XXZ}
\end{equation}
The only non-trivial diagonalization occurs in the $q=2$ block. Restricting to this sector, we now invoke the translation symmetry of the model
\begin{equation}
    \Gamma_{4} H^{2}_{\rm XXZ}\Gamma_{4}^{\dag}=H^{2,0}_{\rm XXZ}\oplus H^{2,1}_{\rm XXZ}\oplus H^{2,2}_{\rm XXZ}\oplus H^{2,3}_{\rm XXZ}
\end{equation}
where $\Gamma_{4}$ is a unitary transformation that diagonalizes $T_{4}$. The $a=1,3$ blocks are singlets, while $a=0,2$ are doublets (see Eq.(\ref{eq:4Qextradegenerate})). Only for $a=0$ do the Fourier modes mix, although the mixing angle is parameter independent. But obviously the eigenvalues depend on the Hamiltonian paramters. We find the non-unitary parameters of the dynamical map to be 
\begin{equation}
\label{eq:CC}
\begin{split}
    \lambda_{3,i}(t,4) =& \Bigg(\frac{7}{16}{\rm +}\frac{1}{4}\cos{2J_{\perp}t}{\rm +}\frac{5}{16}\cos{4J_{\perp}t}\Bigg) \\ {\rm +}&\Bigg(\frac{1}{16}-\frac{1}{12}\cos{2J_{\perp}t}+\frac{1}{48}\cos{4J_{\perp}t}\Bigg) \sum_{\substack{k<l \\ k,l\neq i}}z_{k}z_{l}
\end{split}
\end{equation}
\begin{equation}
\begin{split}
    \tau_{3,i}(t,4) =& \Bigg(\frac{3}{16}{\rm-}\frac{1}{12}\cos{2J_{\perp}t}{\rm-}\frac{5}{48}\cos{4J_{\perp}t}\Bigg)\sum_{k\neq i}z_{k}\\{\rm-}&\Bigg(\frac{3}{16}{\rm-}\frac{1}{4}\cos{2J_{\perp}t}{\rm+}\frac{1}{16}\cos{4J_{\perp}t}\Bigg)\prod_{k\neq i}z_{k}
\end{split}
\end{equation}

\begin{equation}
    \begin{split}
        \alpha_{i}(t,4)=& \Bigg(\frac{3}{16}\cos{(3J_{\perp}{\rm +}J_{\parallel})t}{\rm +}\frac{1}{16}\cos{3(J_{\perp}{\rm -}J_{\parallel})t}\\{\rm +}&\frac{3}{32}\cos{(J_{\perp}{\rm +}3J_{\parallel})t} {\rm +}\frac{1}{4}\cos{(J_{\perp}+J_{\parallel})t} \\ {\rm+}&\frac{3}{8}\cos{(J_{\perp}{\rm -}J_{\parallel})t}{\rm +}\frac{1}{32}\cos{(5J_{\perp}{\rm -}J_{\parallel})t}\Bigg)\\{\rm +}&\Bigg(\frac{3}{16}\cos{(3J_{\perp}{\rm +}J_{\parallel})t}+\frac{1}{16}\cos{3(J_{\perp}{\rm -}J_{\parallel})t}\\{\rm -}&\frac{1}{32}\cos{(J_{\perp}{\rm +}3J_{\parallel})t} {\rm -}\frac{1}{12}\cos{(J_{\perp}{\rm +}J_{\parallel})t} \\ {\rm -}&\frac{1}{8}\cos{(J_{\perp}{\rm -}J_{\parallel})t}{\rm -}\frac{1}{96}\cos{(5J_{\perp}{\rm -}J_{\parallel})t}\Bigg)\sum_{\substack{k<l \\ k,l\neq i}}z_{k}z_{l}
    \end{split}
\end{equation}

\begin{equation}
\begin{split}
    \beta_{i}(t,4)=&\Bigg(\frac{3}{16}\sin{(3J_{\perp}{\rm +}J_{\parallel})t}{\rm -}\frac{1}{16}\sin{3(J_{\perp}{\rm -}J_{\parallel})t}\\{\rm +}&\frac{1}{32}\sin{(J_{\perp}{\rm +}3J_{\parallel})t} {\rm +}\frac{1}{12}\sin{(J_{\perp}{\rm +}J_{\parallel})t}\\ {\rm-}&\frac{1}{8}\sin{(J_{\perp}{\rm -}J_{\parallel})t}{\rm -}\frac{1}{96}\sin{(5J_{\perp}{\rm -}J_{\parallel})t}\Bigg)\sum_{k\neq i}z_{k}  \\+&\Bigg(\frac{3}{16}\sin{(3J_{\perp}{\rm +}J_{\parallel})t}{\rm -}\frac{1}{16}\sin{3(J_{\perp}{\rm -}J_{\parallel})t} \\{\rm -}&\frac{3}{32}\sin{(J_{\perp}{\rm +}3J_{\parallel})t} {\rm -}\frac{1}{4}\sin{(J_{\perp}{\rm +}J_{\parallel})t}\\{\rm+}&\frac{3}{8}\sin{(J_{\perp}{\rm -}J_{\parallel})t}{\rm +}\frac{1}{32}\sin{(5J_{\perp}{\rm -}J_{\parallel})t} \Bigg)\prod_{k\neq i}z_{k}
\end{split}
\end{equation}
\subsubsection*{5-qubits}
As the exact diagonalization is more complicated for the 5-qubit network, we shall go through the steps more explicitly than the 4-qubit case. Again we may change to the canonical excitation basis
\begin{equation}
\mathcal{P}_{5}H_{\rm XXZ}\mathcal{P}_{5}^{\dag}= H^{0}_{\rm XXZ}\oplus H^{1}_{\rm XXZ}\oplus H^{2}_{\rm XXZ}\oplus H^{3}_{\rm XXZ} \oplus H^{4}_{\rm XXZ}\oplus H^{5}_{\rm XXZ} \,.
\end{equation}
As with the ring topology, we further go to the basis of translation eigenstates. Again the only excitation number blocks that are not diagonalized trivially at this stage are
\begin{equation}
\Gamma_{5}H^2_{\rm XXZ}\oplus H^3_{\rm XXZ}\Gamma^{\dag}_{5}=\bigoplus _{a=0}^{4} H^{2,a}_{\rm XXZ}\oplus H^{3,a}_{\rm XXZ} \,.
\end{equation}
And as for the ring-topology, we obtain a set of ten 2$\times$2 matrices that require further computation to diagonalize. The $q=2$ blocks are found to be 
\begin{equation}
\begin{split}
H^{2,0}_{\rm XXZ}&=
\begin{bmatrix}
-h-J_{\parallel}+2J_{\perp}&4J_{\perp}\\ \\ 4J_{\perp}& -h-J_{\parallel}+2J_{\perp}
    \end{bmatrix} \\
H^{2,1}_{\rm XXZ}=&\begin{bmatrix}
        -h-J_{\parallel}+\left(\frac{\sqrt{5}-1}{2}\right)J_{\perp}&-J_{\perp}e^{-\frac{6\pi i }{5}}\\ \\ -J_{\perp}e^{\frac{6\pi i }{5}}& -h-J_{\parallel}-\left(\frac{\sqrt{5}+1}{2}\right)J_{\perp}
    \end{bmatrix} \\
      H^{2,2}_{\rm XXZ}=&\begin{bmatrix}
        -h-J_{\parallel}-\left(\frac{\sqrt{5}+1}{2}\right)J_{\perp}&-J_{\perp}e^{-\frac{2\pi i}{5}}\\ \\ -J_{\perp}e^{\frac{2\pi i}{5}}& -h-J_{\parallel}+\left(\frac{\sqrt{5}-1}{2}\right)J_{\perp}
    \end{bmatrix}
    \end{split}
\end{equation}
With the remaining blocks determined by the relations $H^{2,2}_{\rm XXZ}=\left(H^{2,3}_{\rm XXZ}\right)^{T}$ and $H^{2,1}_{\rm XXZ}=\left(H^{2,4}_{\rm XXZ}\right)^{T}$. 
The matrix representations for the $q=3$ block are obtained through the replacement $h\rightarrow -h$, affecting only the eigenvalues and not the form of the associated eigenvectors. Again due to large degree of symmetry of the model, the eigenvectors are parameter independent. The resulting map parameter functions are:
\begin{equation}
\begin{split}
    \lambda_{3,i}(t,5) =&\Bigg(\frac{7}{15}{\rm +}\frac{1}{3}\cos{3J_{\perp}t}{\rm +}\frac{1}{5}\cos{5J_{\perp}t}\Bigg) \\{\rm +}&\Bigg(\frac{8}{225}-\frac{1}{18}\cos{3J_{\perp}t}+\frac{1}{50}\cos{5J_{\perp}t}\Bigg)\sum_{\substack{k<l \\ k,l\neq i}}z_{k}z_{l}
\end{split}
\end{equation}
\begin{equation}
\begin{split}
    \tau_{3,i}(t,5) =&\Bigg(\frac{2}{15}-\frac{1}{12}\cos{3J_{\perp}t}-\frac{1}{20}\cos{5J_{\perp}t}\Bigg)\sum_{k\neq i}z_{k}\\-&\Bigg(\frac{4}{75}-\frac{1}{12}\cos{3J_{\perp}t}+\frac{3}{100}\cos{5J_{\perp}t}\Bigg)\smashoperator{\sum_{\substack{k<l<j \\ k,l,j\neq i}}}z_{k}z_{l}z_{j}
\end{split}
\end{equation}
\begin{equation}
    \begin{split}
        \alpha_{i}(t,5)=&\Bigg(\frac{157}{600}+\frac{1}{12}\cos{3J_{\perp}t}+\frac{3}{100}\cos{5J_{\perp}t}+\frac{3}{50}\cos{(3J_{\perp}+2J_{\parallel})t}\\+&\frac{1}{40}\cos{4(J_{\perp}{\rm-}J_{\parallel})t}+\frac{1}{100}\cos{(7J_{\perp}-2J_{\parallel})t}\\+&\frac{9}{50}\cos{2(J_{\perp}-J_{\parallel})t}+\frac{1}{4}\cos{(J_{\perp}+2J_{\parallel})t}\\{\rm +}&\frac{1}{10}\cos{(J_{\perp}{\rm+}4J_{\parallel})t}\Bigg) \\ -&\Bigg(\frac{157}{1800}+\frac{1}{36}\cos{3J_{\perp}t}+\frac{1}{100}\cos{5J_{\perp}t}-\frac{1}{40}\cos{4(J_{\perp}-J_{\parallel})t}\\-&\frac{1}{10}\cos{(J_{\perp} +4J_{\parallel})t}\Bigg)\sum_{\substack{k<l \\ k,l\neq i}}z_{k}z_{l} \\ +&\Bigg(\frac{157}{600}+\frac{1}{12}\cos{3J_{\perp}t}+\frac{3}{100}\cos{5J_{\perp}t}+\frac{1}{10}\cos{(J_{\perp}+4J_{\parallel})t}\\-&\frac{1}{4}\cos{(J_{\perp}+2J_{\parallel})t}-\frac{3}{50}\cos{(3J_{\perp}+2J_{\parallel})t}\\-&\frac{1}{100}\cos{(7J_{\perp}-2J_{\parallel})t}-\frac{9}{50}\cos{2(J_{\perp}-J_{\parallel})t}\\+&\frac{1}{40}\cos{4(J_{\perp}-J_{\parallel})t}\Bigg)\,\,\,\smashoperator{\sum_{\substack{j<k<l<m \\ j,k,l,m\neq i}}}z_{j}z_{k}z_{l}z_{m}
    \end{split}
\end{equation}
\begin{equation}
\begin{split}
    \beta_{i}(t,5)=&\Bigg(\frac{3}{100}\sin{(3J_{\perp}+2J_{\parallel})t}-\frac{1}{40}\sin{4(J_{\perp}-J_{\parallel})t}\\-&\frac{1}{200}\sin{(7J_{\perp}-2J_{\parallel})t}-\frac{9}{100}\sin{2(J_{\perp}-J_{\parallel})t}\\+&\frac{1}{8}\sin{(J_{\perp}+2J_{\parallel})t}+\frac{1}{10}\sin{(J_{\perp}+4J_{\parallel})t}\Bigg) \sum_{k\neq i}z_{k} \\ -&\Bigg(\frac{3}{100}\sin{(3J_{\perp}+2J_{\parallel})t}+\frac{1}{40}\sin{4(J_{\perp}-J_{\parallel})t}\\-&\frac{1}{200}\sin{(7J_{\perp}-2J_{\parallel})t}-\frac{9}{100}\sin{2(J_{\perp}-J_{\parallel})t}\\+&\frac{1}{8}\sin{(J_{\perp}+2J_{\parallel})t}-\frac{1}{10}\sin{(J_{\perp}+4J_{\parallel})t}\Bigg)\smashoperator{\sum_{\substack{k<l<j \\ j,k,l\neq i}}}z_{j}z_{k}z_{l} 
\end{split}
\end{equation}

\subsubsection*{6-qubits}
Repeating the procedures of the previous sections we may block diagonalize the 6-qubit XXZ-network. Going to the Fourier basis we have the remaining blocks to diagonalize
\begin{equation}
    \Gamma_{6}\Big(H^{2}_{\rm XXZ}\oplus H^{3}_{\rm XXZ}\oplus H^{4}_{\rm XXZ} \Big)\Gamma^{\dag}_{6}=\bigoplus_{a=0}^{5}H^{2,a}_{\rm XXZ}\oplus H^{3,a}_{\rm XXZ} \oplus H^{4,a}_{\rm XXZ}
\end{equation}
which provides more of a challenge than the previous examples as now $4\times 4$ blocks and $3\times 3$ blocks appear. But the example still retains a degree of simplicity, as the eigenvectors do not contain any parameter dependence. The computation may be further reduced by noting that once $H^{2,a}_{\rm XXZ}$ is diagonalized, a similar transformation may be used to diagonalize $H^{4,a}_{\rm XXZ}$. As for the size of the various blocks we have for $q=2$
\begin{equation}
    \begin{split}
        &{\rm dim}H^{2,0}_{\rm XXZ}={\rm dim}H^{2,2}_{\rm XXZ}={\rm dim}H^{2,4}_{\rm XXZ}=3 \\
        &{\rm dim}H^{2,1}_{\rm XXZ}={\rm dim}H^{2,3}_{\rm XXZ}={\rm dim}H^{2,5}_{\rm XXZ}=2
    \end{split}
\end{equation}
where the same holds for $q=4$. All of these blocks create superpositions amongst the Fourier modes except $H^{2,3}_{\rm XXZ}$, which is comprised of two singlet states $|\mathcal{F}^{3}_{2},0\rangle$ and $|\mathcal{F}^{3}_{2},1\rangle$. The same conclusion applies for $q=4$. For $q=3$ we have
\begin{equation}
   \begin{split}
        &{\rm dim}H^{3,0}_{\rm XXZ}={\rm dim}H^{3,3}_{\rm XXZ}=4 \\
         &{\rm dim}H^{3,1}_{\rm XXZ}={\rm dim}H^{3,2}_{\rm XXZ}={\rm dim}H^{3,4}_{\rm XXZ}={\rm dim}H^{3,5}_{\rm XXZ}=3 
   \end{split} 
\end{equation}
and there are mixtures created among all the Fourier modes except $|\mathcal{F}^{2}_{3},0\rangle$ and $|\mathcal{F}^{4}_{3},0\rangle$.
\begin{equation}
\label{eq:CC1}
\begin{split}
    \lambda_{3,i}(t,6) &=\Bigg(\frac{59}{144}{\rm +}\frac{5}{32}\cos{2J_{\perp}t}{\rm +}\frac{5}{16}\cos{4J_{\perp}t}+\frac{35}{288}\cos{6J_{\perp}t}\Bigg) \\&+\Bigg(\frac{1}{24}{\rm -}\frac{1}{32}\cos{2J_{\perp}t}{\rm -}\frac{1}{40}\cos{4J_{\perp}t}+\frac{7}{480}\cos{6J_{\perp}t}\Bigg)\sum_{\substack{k<l \\ k,l\neq i}}z_{k}z_{l}
\\&-\Bigg(\frac{1}{48}{\rm -}\frac{1}{32}\cos{2J_{\perp}t}{\rm +}\frac{1}{80}\cos{4J_{\perp}t}-\frac{1}{480}\cos{6J_{\perp}t}\Bigg)\,\,\,\smashoperator{\sum_{\substack{j<k<l,m \\j, k,l,m\neq i}}}z_{j}z_{k}z_{l}z_{m}
\end{split}
\end{equation}
\begin{equation}
\label{eq:CC2}
\begin{split}
    \tau_{3,i}(t,6) =&\Bigg(\frac{17}{144}{\rm -}\frac{1}{32}\cos{2J_{\perp}t}{\rm -}\frac{1}{16}\cos{4J_{\perp}t}-\frac{7}{288}\cos{6J_{\perp}t}\Bigg)\sum_{k\neq i}z_{k} \\-&\Bigg(\frac{1}{24}{\rm -}\frac{1}{32}\cos{2J_{\perp}t}{\rm -}\frac{1}{40}\cos{4J_{\perp}t}+\frac{7}{480}\cos{6J_{\perp}t}\Bigg)\smashoperator{\sum_{\substack{k<l<j \\ k,l,j\neq i}}}z_{k}z_{l}z_{j}
    \\+&\Bigg(\frac{5}{48}{\rm -}\frac{5}{32}\cos{2J_{\perp}t}{\rm +}\frac{1}{16}\cos{4J_{\perp}t}-\frac{1}{96}\cos{6J_{\perp}t}\Bigg)\prod_{j\neq i}z_{j}
\end{split}
\end{equation}
The expressions for $\alpha(t)$ and $\beta(t)$ are lengthy, therefore we shall express our results using the partial components. The remaining non-zero partial components are obtained using the permutation symmetry, for example $\alpha^{zz000}_{i}=\alpha^{z0z00}_{i}=\alpha^{00zz0}_{i}$, etc. We find 
\begin{equation}
\label{eq:CC3}
    \begin{split}
        \alpha^{00000}_{i}(t,6)=&\frac{5}{96}\cos{(J_{\perp}+5J_{\parallel})t}+\frac{1}{96}\cos{5(J_{\perp}-J_{\parallel})t}\\+&\frac{3}{16}\cos{(J_{\perp}+3J_{\parallel})t}+\frac{25}{288}\cos{3(J_{\perp}-J_{\parallel})t}\\+&\frac{1}{288}\cos{(9J_{\perp}-3J_{\parallel})t}+\frac{1}{48}\cos{(5J_{\perp}+J_{\parallel})t}\\+&\frac{3}{32}\cos{(3J_{\perp}+J_{\parallel})t}+\frac{5}{32}\cos{(J_{\perp}+J_{\parallel})t}\\+&\frac{5}{16}\cos{(J_{\perp}-J_{\parallel})t}+\frac{1}{32}\cos{(5J_{\perp}-J_{\parallel})t}\\+&\frac{1}{96}\cos{(7J_{\perp}-J_{\parallel})t}+\frac{5}{144}\cos{3(J_{\perp}+J_{\parallel})t}
      \end{split}
\end{equation}
\begin{equation}
\label{eq:CC4}
    \begin{split}  
    \alpha^{zz000}_{i}(t,6)=&\frac{5}{96}\cos{(J_{\perp}+5J_{\parallel})t}+\frac{1}{96}\cos{5(J_{\perp}-J_{\parallel})t}\\+&\frac{3}{80}\cos{(J_{\perp}+3J_{\parallel})t}-\frac{1}{240}\cos{(5J_{\perp}+J_{\parallel})t}\\-&\frac{3}{160}\cos{(3J_{\perp}+J_{\parallel})t}-\frac{1}{32}\cos{(J_{\perp}+J_{\parallel})t}\\+&\frac{5}{288}\cos{3(J_{\perp}-J_{\parallel})t}+\frac{1}{1440}\cos{(9J_{\perp}-3J_{\parallel})t}\\-&\frac{1}{16}\cos{(J_{\perp}-J_{\parallel})t}-\frac{1}{160}\cos{(5J_{\perp}-J_{\parallel})t}\\-&\frac{1}{480}\cos{(7J_{\perp}-J_{\parallel})t}+\frac{1}{144}\cos{3(J_{\perp}+J_{\parallel})t}
      \end{split}
\end{equation}
\begin{equation}
\label{eq:CC5}
    \begin{split}  
      \alpha^{zzzz0}_{i}(t,6)=&\frac{5}{96}\cos{(J_{\perp}+5J_{\parallel})t}+\frac{1}{96}\cos{5(J_{\perp}-J_{\parallel})t}\\-&\frac{9}{80}\cos{(J_{\perp}+3J_{\parallel})t}-\frac{5}{96}\cos{3(J_{\perp}-J_{\parallel})t}\\-&\frac{1}{480}\cos{(9J_{\perp}-3J_{\parallel})t}+\frac{1}{240}\cos{(5J_{\perp}+J_{\parallel})t}\\+&\frac{3}{160}\cos{(3J_{\perp}+J_{\parallel})t}+\frac{1}{32}\cos{(J_{\perp}+J_{\parallel})t}\\+&\frac{1}{16}\cos{(J_{\perp}-J_{\parallel})t}+\frac{1}{160}\cos{(5J_{\perp}-J_{\parallel})t}\\+&\frac{1}{480}\cos{(7J_{\perp}-J_{\parallel})t}-\frac{1}{48}\cos{3(J_{\perp}+J_{\parallel})t}
\end{split}
\end{equation}
\begin{equation}
\label{eq:CC6}
\begin{split}
    \beta^{z0000}_{i}(t,6)=&\frac{5}{96}\sin{(J_{\perp}+5J_{\parallel})t}-\frac{1}{96}\sin{5(J_{\perp}-J_{\parallel})t}\\+&\frac{9}{80}\sin{(J_{\perp}+3J_{\parallel})t}-\frac{5}{96}\sin{3(J_{\perp}-J_{\parallel})t}\\-&\frac{1}{480}\sin{(9J_{\perp}-3J_{\parallel})t}+\frac{1}{240}\sin{(5J_{\perp}+J_{\parallel})t}\\+&\frac{3}{160}\sin{(3J_{\perp}+J_{\parallel})t}+\frac{1}{32}\sin{(J_{\perp}+J_{\parallel})t}\\-&\frac{1}{16}\sin{(J_{\perp}-J_{\parallel})t}-\frac{1}{160}\sin{(5J_{\perp}-J_{\parallel})t}\\-&\frac{1}{480}\sin{(7J_{\perp}-J_{\parallel})t}+\frac{1}{48}\sin{3(J_{\perp}+J_{\parallel})t} 
 \end{split}
\end{equation}
\begin{equation}
\label{eq:CC7}
    \begin{split}   
    \beta^{zzz00}_{i}(t,6)=&\frac{5}{96}\sin{(J_{\perp}+5J_{\parallel})t}-\frac{1}{96}\sin{5(J_{\perp}-J_{\parallel})t}\\-&\frac{3}{80}\sin{(J_{\perp}+3J_{\parallel})t}+\frac{5}{288}\sin{3(J_{\perp}-J_{\parallel})t}\\+&\frac{1}{1440}\sin{(9J_{\perp}-3J_{\parallel})t}-\frac{1}{240}\sin{(5J_{\perp}+J_{\parallel})t}\\-&\frac{3}{160}\sin{(3J_{\perp}+J_{\parallel})t}-\frac{1}{32}\sin{(J_{\perp}+J_{\parallel})t}\\+&\frac{1}{16}\sin{(J_{\perp}-J_{\parallel})t}+\frac{1}{160}\sin{(5J_{\perp}-J_{\parallel})t}\\+&\frac{1}{480}\sin{(7J_{\perp}-J_{\parallel})t}-\frac{1}{144}\sin{3(J_{\perp}+J_{\parallel})t} 
\end{split}
\end{equation}
\begin{equation}
\label{eq:CC8}
    \begin{split}
    \beta^{zzzzz}_{i}(t,6)=&\frac{5}{96}\sin{(J_{\perp}+5J_{\parallel})t}-\frac{1}{96}\sin{5(J_{\perp}-J_{\parallel})t}\\-&\frac{3}{16}\sin{(J_{\perp}+3J_{\parallel})t}+\frac{25}{288}\sin{3(J_{\perp}-J_{\parallel})t}\\+&\frac{1}{288}\sin{(9J_{\perp}-3J_{\parallel})t}+\frac{1}{48}\sin{(5J_{\perp}+J_{\parallel})t}\\+&\frac{3}{32}\sin{(3J_{\perp}+J_{\parallel})t}+\frac{5}{32}\sin{(J_{\perp}+J_{\parallel})t}\\-&\frac{5}{16}\sin{(J_{\perp}-J_{\parallel})t}-\frac{1}{32}\sin{(5J_{\perp}-J_{\parallel})t}\\-&\frac{1}{96}\sin{(7J_{\perp}-J_{\parallel})t}-\frac{5}{144}\sin{3(J_{\perp}+J_{\parallel})t}
\end{split}
\end{equation}
\subsection*{Ring-connected XXZ-network}
\label{RCXXZ}
Turning to the more difficult model, the homogeneous XXZ-ring. Due to the more complicated structure of this model we only investigate the XXZ model for $N=4$, and use the XXX model for $N=5$, where we only present the computations for $\lambda_{3}(t)$ and $\tau_{3}(t)$. 
\subsubsection*{4-qubits}
For 4-qubits the Hamiltonian has the block diagonal structure
\begin{equation}
\mathcal{P}_{4}H_{\rm XXZ}\mathcal{P}_{4}^{\dag}= H^{0}_{\rm XXZ}\oplus H^{1}_{\rm XXZ}\oplus H^{2}_{\rm XXZ}\oplus H^{3}_{\rm XXZ} \oplus H^{4}_{\rm XXZ}
\end{equation}
 The only non-trivial diagonalization occurs in the $q$=2 block. Restricting to this sector, we may reduce it to smaller blocks by invoking the translation symmetry
\begin{equation}
    \Gamma_{4} H^{2}_{\rm XXZ}\Gamma_{4}^{\dag}=H^{2,0}_{\rm XXZ}\oplus H^{2,1}_{\rm XXZ}\oplus H^{2,2}_{\rm XXZ}\oplus H^{2,3}_{\rm XXZ}
\end{equation}
where $\Gamma_{4}$ is a unitary transformation that diagonalizes $T_{4}$. The $a=1,3$ blocks are singlets, while $a=0,2$ are both doublets. Mixing only occurs within the $a=0$ block, which in this case does depend on the parameters of the Hamiltonian.  
Theses stationary states of $H^{2,2}_{\rm XXZ}$ are determined to be 
\begin{equation}
\begin{split}
&|E^{0}_{2};0\rangle=\cos\frac{\phi_{\times}}{2}|\mathcal{F}^{0}_{2};0\rangle+\sin\frac{\phi_{\times}}{2}|\mathcal{F}^{0}_{2};1\rangle \\
&|E^{0}_{2};1\rangle=\sin\frac{\phi_{\times}}{2}|\mathcal{F}^{0}_{2};0\rangle-\cos\frac{\phi_{\times}}{2}|\mathcal{F}^{0}_{2};1\rangle 
\\&\omega_{\pm}=J_{\parallel}\pm{\rm sgn}\big(J_{\parallel}\big) J_{\times}
\end{split}
\end{equation}

where the Hamiltonian dependent parameters we have introduced are
\begin{equation}
\begin{split}
&J_{\times}=\sqrt{J_{\parallel}^{2}+8J_{\perp}^{2}} \\
&\phi_{\times}=-{\rm tan}^{-1}2^{\frac{3}{2}}\frac{J_{\perp}}{J_{\parallel}} \,. 
\end{split}
\end{equation}

 \begin{equation}
 \label{eq:l34qubitting}
 \begin{split}
    \lambda_{3,i}(t,4) &=\Big(\frac{7}{16}+\frac{1}{4}\cos2J_{\perp}t+\frac{1}{16}\cos4J_{\perp}t+\frac{1}{4}\cos J_{\parallel}t\cos J_{\times}t\\&+\frac{1}{4}\cos\phi_{\times}\sin J_{\parallel}t\sin J_{\times}t\Big)\\ &+\Big(\frac{1}{16}+\cos4J_{\perp}t-\frac{1}{8}\cos J_{\parallel}t\cos J_{\times}t\\&-\frac{1}{8}\cos\phi_{\times}\sin J_{\parallel}t\sin J_{\times}t \Big)\bigg(z_{i-1}z_{i+2}+z_{i+1}z_{i+2}\bigg)\\ &-\Big(\frac{3}{16}-\frac{1}{4}\cos2J_{\perp}t+\frac{1}{16}\cos4J_{\perp}t\Big)\bigg(z_{i-1}z_{i+1}\bigg)
\end{split}
\end{equation}
\begin{equation}
\begin{split}
    \tau_{3,i}(t,4) &=\Big(\frac{1}{16}-\frac{1}{4}\cos2J_{\perp}t-\frac{1}{16}\cos4J_{\perp}t+\frac{1}{4}\cos J_{\parallel}t\cos J_{\times}t\\&+\frac{1}{4}\cos\phi_{\times}\sin J_{\parallel}t\sin J_{\times}t\Big)\prod_{k \neq i}z_{k} \\&+\Big(\frac{3}{16}-\frac{1}{16}\cos4J_{\perp}t-\frac{1}{8}\cos\phi_{\times}\sin J_{\parallel}t\sin J_{\times}t\\&-\frac{1}{8}\cos J_{\parallel}t\cos J_{\times}t\Big) (z_{i-1}+z_{i+1})\\ &+\Big(\frac{3}{16}-\frac{1}{4}\cos2J_{\perp}t+\frac{1}{16}\cos4J_{\perp}t\Big)z_{i+2}
\end{split}
\end{equation}
As seen in the previous section, the expression for $\alpha(t)$ and $\beta(t)$ are more cumbersome than the other non-unitary parameters. Therefore we again split them into their partial components. For the ring topology we only need to consider the reflection symmetry to generate additional partial components as will be clear in the following equations. We find
\begin{equation}
    \begin{split}
        \alpha^{000}_{i}(t,4)&=\frac{1}{4} {\rm+}\frac{1{\rm+}3\cos2J_{\parallel}t}{16}\cos^{2}J_{\perp}t{\rm +}\frac{1{\rm+}3\cos2J_{\perp}t}{16}\cos J_{\parallel}t\cos J_{\times}t \\ &-\frac{\sqrt{2}}{8}\sin\phi_{\times}\sin2J_{\perp}t\cos J_{\parallel}t\sin J_{\times}t\\&{\rm-}\frac{\big(1{\rm+}\cos\phi_{\times}\sin J_{\perp}t\sin J_{\times}t\big)}{8}\sin^{2}J_{\parallel}t
        \end{split}
\end{equation} 
\begin{equation}
    \begin{split}
         \alpha^{0zz}_{i}(t,4)&=\frac{\cos^{2}{J_{\perp}}t}{8}\big(\cos2J_{\parallel}t{\rm-}\cos J_{\parallel}t\cos J_{\times}t{\rm +}\cos\phi_{\times}\sin J_{\parallel}t\sin J_{\times}t\big) \\&{\rm+}\frac{\sqrt{2}}{16}\sin\phi_{\times}\cos J_{\parallel}t\sin2J_{\perp}t\sin J_{\times}t= \alpha^{zz0}_{i}(t,4)
\end{split}
\end{equation}

\begin{equation}
    \begin{split}      
        \alpha^{z0z}_{i}(t,4)=&{\rm-}\frac{1}{8}+\frac{\cos^{2}J_{\perp}t}{4}\big(1{\rm+}\frac{3}{2}\cos2J_{\parallel}t\big){\rm-}\frac{\sin^{2}J_{\perp}t}{8}\cos J_{\parallel}t\cos J_{\times}t\\&{\rm+}\frac{1+3\cos2J_{\perp}t}{16}\cos\phi_{\times}\sin J_{\parallel}t\sin J_{\times}t
\end{split}
\end{equation}
\begin{equation}
\begin{split}
        \beta^{z00}_{i}(t,4)=&{\rm-}\frac{\cos^{2}{J_{\perp}t}}{8}\Big(3\sin2J_{\parallel}t{\rm+}\sin J_{\parallel}t\cos J_{\times}t{\rm+}\cos\phi_{\times}\cos J_{\parallel}t\sin J_{\times}t\Big)\\&+\frac{\sqrt{2}}{16}\sin\phi_{\times}\sin J_{\parallel}t\sin2J_{\perp}t\sin J_{\times}t=\beta^{00z}_{i}(t,4)
      \end{split}
\end{equation}
\begin{equation}
\begin{split}  
     \beta^{0z0}_{i}(t,4)&=\frac{1{\rm+}3\cos2J_{\perp}t}{16}\cos\phi_{\times}\cos J_{\parallel}t\sin J_{\times}t{\rm-}\frac{\cos^{2}J_{\perp}t}{8}\sin2J_{\parallel}t\\&{\rm+}\frac{\sin^{2}J_{\perp}t}{8}\sin J_{\parallel}t\cos J_{\times}t
\end{split}
\end{equation}
\begin{equation}
\begin{split}    
   \beta^{zzz}_{i}(t,4)=&\frac{\sin^{2}J_{\perp}t}{8}\cos\phi_{\times}t\cos J_{\parallel}t\sin J_{\times}t-\frac{\cos^{2}J_{\perp}t}{8}\sin2J_{\parallel}t\\+&\frac{1+3\cos2J_{\perp}t}{16}\sin J_{\parallel}t\cos J_{\times}t\\-&\frac{\sqrt{2}}{16}\sin\phi_{\times}\sin J_{\parallel}t\sin2J_{\perp}t\sin J_{\times}t
    \end{split}
\end{equation}

\subsubsection*{5-qubits}
As with the previous examples we change from the computational basis to the canonically  ordered excitation basis. We achieve this by using a permutation matrix
\begin{equation}
\mathcal{P}_{5}H_{\rm XXZ}\mathcal{P}_{5}^{\dag}= H^{0}_{\rm XXZ}\oplus H^{1}_{\rm XXZ}\oplus H^{2}_{\rm XXZ}\oplus H^{3}_{\rm XXZ} \oplus H^{4}_{\rm XXZ}\oplus H^{5}_{\rm XXZ}
\end{equation}
and we are interested in finding the stationary states of the middle two blocks, as the others may be found trivially. Now we can go further by changing to a translation eigenbasis
\begin{equation}
\Gamma_{5}H^2_{\rm XXZ}\oplus H^3_{\rm XXZ}\Gamma^{\dag}_{5}=\bigoplus _{a=0}^{4} H^{2,a}_{\rm XXZ}\oplus H^{3,a}_{\rm XXZ}
\end{equation}
where the $q=2$ blocks take the form
\begin{equation}
    H^{2,a}_{\rm XXZ}=\begin{bmatrix}
        -h+\frac{1}{2}J_{\parallel}&\big(2\cos\frac{6\pi a}{5}J_{\perp}\big) e^{-\frac{6\pi i a}{5}}\\ \\ \big(2\cos\frac{6\pi a}{5}J_{\perp}\big)e^{\frac{6\pi i a}{5}}& -h-\frac{3}{2}J_{\parallel}+2\cos{\frac{6\pi a}{5}J_{\perp}}
    \end{bmatrix}
\end{equation}
and the $q=3$ blocks are obtained by sending $h\rightarrow -h $. We thus obtain the following parameterization of the non-trivial stationary states
\begin{equation}
\begin{split}   &\tan\psi_{a}=\frac{6\pi a}{5} \\
&\tan\phi_{a}=\frac{2\cos{\frac{6\pi a}{5}}J_{\perp}}{\big(J_{\parallel}-\cos{\frac{6\pi a}{5}}J_{\perp}\big)}\\
 & J_{a}=\sqrt{(J_{\parallel}-\cos{\frac{6\pi a}{5}}J_{\perp})^{2}+4\cos^{2}{\frac{6\pi a}{5}}J^{2}_{\perp}}
 \end{split}
\end{equation}
Due to the length of the expression obtained, we must report the non-unitary parameters by the partial components defined in the main text. Setting $J_{\perp}=J_{\parallel}$ we find
 \begin{equation}
 \label{eq:l35qubitring}
     \begin{split}
         \lambda^{0000}_{3,i}(t;5)&=\frac{71}{225}+\frac{13}{90}\cos{(\frac{3-\sqrt{5}}{2})J_{\perp}t}+\frac{1}{10}\cos{(\frac{5-\sqrt{5}}{2})J_{\perp}t}\\&+\frac{1}{45}\cos{3(\frac{1-\sqrt{5}}{2})J_{\perp}t}+\frac{32}{225}\cos{\sqrt{5}J_{\perp}t}\\&+\frac{13}{90}\cos{(\frac{3+\sqrt{5}}{2})J_{\perp}t}+\frac{1}{10}\cos{(\frac{5+\sqrt{5}}{2})J_{\perp}t}\\&+\frac{2}{225}\cos{2\sqrt{5}J_{\perp}t}+\frac{1}{45}\cos{3(\frac{1+\sqrt{5}}{2})J_{\perp}t}
    \end{split}
    \end{equation}
    \begin{equation}
    \begin{split}
        \lambda^{zz00}_{3,i}(t,5)=&-\frac{6\sqrt{5}{\rm+}25}{900}\cos{(\frac{3{\rm+}\sqrt{5}}{2})J_{\perp}t}{\rm+}\frac{6\sqrt{5}{\rm-}25}{900}\cos{(\frac{3{\rm-}\sqrt{5}}{2})J_{\perp}t}\\+&\frac{1}{100}\cos{(\frac{5+\sqrt{5}}{2})J_{\perp}t}+\frac{1}{100}\cos{(\frac{5-\sqrt{5}}{2})J_{\perp}t}\\-&\frac{\sqrt{5}}{450}\cos{3(\frac{1-\sqrt{5}}{2})J_{\perp}t}+\frac{\sqrt{5}}{450}\cos{3(\frac{1+\sqrt{5}}{2})J_{\perp}t}\\-&\frac{2}{225}\cos{2\sqrt{5}J_{\perp}t}+\frac{2}{45}\cos{\sqrt{5}J_{\perp}t}= \lambda^{00zz}_{3,i}(t,5)
    \end{split}
    \end{equation}    
    \begin{equation}
    \begin{split}
        \lambda^{z0z0}_{3,i}(t,5)=&\frac{1}{100}\cos{(\frac{5-\sqrt{5}}{2})J_{\perp}t}-\frac{6\sqrt{5}+25}{900}\cos{(\frac{3-\sqrt{5}}{2})J_{\perp}t}\\&+\frac{\sqrt{5}}{450}\cos{3(\frac{1-\sqrt{5}}{2})J_{\perp}t}-\frac{\sqrt{5}}{450}\cos{3(\frac{1+\sqrt{5}}{2})J_{\perp}t}\\&+\frac{6\sqrt{5}-25}{900}\cos{(\frac{3+\sqrt{5}}{2})J_{\perp}t}+\frac{1}{100}\cos{(\frac{5+\sqrt{5}}{2})J_{\perp}t}\\&-\frac{2}{225}\cos{2\sqrt{5}J_{\perp}t}+\frac{2}{45}\cos{\sqrt{5}J_{\perp}t}= \lambda^{0z0z}_{3,i}(t,5)
    \end{split}
    \end{equation}     
     \begin{equation}
     \begin{split}
         \lambda^{0zz0}_{3,i}(t,5)&=\frac{1}{45}+\frac{3\sqrt{5}-5}{300}\cos{(\frac{3-\sqrt{5}}{2})J_{\perp}t}+\frac{1}{75}\cos{2\sqrt{5}J_{\perp}t}\\&+\frac{1-\sqrt{5}}{100}\cos{(\frac{5-\sqrt{5}}{2})J_{\perp}t}-\frac{5-\sqrt{5}}{450}\cos{3(\frac{1-\sqrt{5}}{2})J_{\perp}t}\\&-\frac{3\sqrt{5}+5}{300}\cos{(\frac{3+\sqrt{5}}{2})J_{\perp}t}+\frac{1+\sqrt{5}}{100}\cos{(\frac{5+\sqrt{5}}{2})J_{\perp}t}\\&-\frac{5+\sqrt{5}}{450}\cos{3(\frac{1+\sqrt{5}}{2})J_{\perp}t}
    \end{split}
    \end{equation}
    \begin{equation}
    \begin{split}
         \lambda^{z00z}_{3,i}(t,5)&=\frac{1}{45}{\rm-}\frac{3\sqrt{5}{\rm+}5}{300}\cos{(\frac{3{\rm-}\sqrt{5}}{2})J_{\perp}t}{\rm+}\frac{1{\rm+}\sqrt{5}}{100}\cos{(\frac{5{\rm-}\sqrt{5}}{2})J_{\perp}t}\\&-\frac{5+\sqrt{5}}{450}\cos{3(\frac{1-\sqrt{5}}{2})J_{\perp}t}+\frac{3\sqrt{5}-5}{300}\cos{(\frac{3+\sqrt{5}}{2})J_{\perp}t}\\&+\frac{1-\sqrt{5}}{100}\cos{(\frac{5+\sqrt{5}}{2})J_{\perp}t}+\frac{1}{75}\cos{2\sqrt{5}J_{\perp}t}\\&+\frac{5-\sqrt{5}}{450}\cos{3(\frac{1+\sqrt{5}}{2})J_{\perp}t}
     \end{split}
 \end{equation}
 \begin{equation}
     \begin{split}
         \tau^{z000}_{3,i}(t;5)&=\frac{77}{450}{\rm-}\frac{1{\rm+}\sqrt{5}}{40}\cos{(\frac{5{\rm+}\sqrt{5}}{2})J_{\perp}t}{\rm-}\frac{1{\rm-}\sqrt{5}}{40}\cos{(\frac{5{\rm-}\sqrt{5}}{2})J_{\perp}t}\\&-\frac{1-\sqrt{5}}{180}\cos{3(\frac{1-\sqrt{5}}{2})J_{\perp}t}-\frac{1+\sqrt{5}}{180}\cos{3(\frac{1+\sqrt{5}}{2})J_{\perp}t}\\&-\frac{13+3\sqrt{5}}{360}\cos{(\frac{3+\sqrt{5}}{2})J_{\perp}t}{\rm-}\frac{13{\rm-}3\sqrt{5}}{360}\cos{(\frac{3{\rm-}\sqrt{5}}{2})J_{\perp}t}\\&-\frac{1}{450}\cos{2\sqrt{5}J_{\perp}t}-\frac{8}{225}\cos{\sqrt{5}J_{\perp}t}=\tau^{000z}_{3,i}(t;5)
      \end{split}
 \end{equation}
 \begin{equation}
     \begin{split}    
         \tau^{0z00}_{3,i}(t;5)&=\frac{77}{450}{\rm-}\frac{1{\rm-}\sqrt{5}}{40}\cos{(\frac{5{\rm+}\sqrt{5}}{2})J_{\perp}t}{\rm-}\frac{1{\rm+}\sqrt{5}}{40}\cos{(\frac{5{\rm-}\sqrt{5}}{2})J_{\perp}t}\\&-\frac{1{\rm+}\sqrt{5}}{180}\cos{3(\frac{1{\rm-}\sqrt{5}}{2})J_{\perp}t}{\rm-}\frac{1{\rm-}\sqrt{5}}{180}\cos{3(\frac{1{\rm+}\sqrt{5}}{2})J_{\perp}t}\\&-\frac{13{\rm-}3\sqrt{5}}{360}\cos{(\frac{3{\rm+}\sqrt{5}}{2})J_{\perp}t}{\rm-}\frac{13{\rm+}3\sqrt{5}}{360}\cos{(\frac{3{\rm-}\sqrt{5}}{2})J_{\perp}t}\\&-\frac{1}{450}\cos{2\sqrt{5}J_{\perp}t}-\frac{8}{225}\cos{\sqrt{5}J_{\perp}t}=\tau^{00z0}_{3,i}(t;5)
       \end{split}
 \end{equation}
 \begin{equation}
     \begin{split}   
          \tau^{zzz0}_{3,i}(t;5)=&{\rm-}\frac{1}{90}{\rm-}\frac{3{\rm-}\sqrt{5}}{200}\cos{(\frac{5{\rm+}\sqrt{5}}{2})J_{\perp}t}{\rm-}\frac{3{\rm+}\sqrt{5}}{200}\cos{(\frac{5{\rm-}\sqrt{5}}{2})J_{\perp}t}\\&{\rm+}\frac{5{\rm+}3\sqrt{5}}{900}\cos{3(\frac{1{\rm-}\sqrt{5}}{2})J_{\perp}t}{\rm+}\frac{5{\rm-}3\sqrt{5}}{900}\cos{3(\frac{1{\rm+}\sqrt{5}}{2})J_{\perp}t}\\&{\rm+}\frac{65{\rm+}3\sqrt{5}}{1800}\cos{(\frac{3{\rm+}\sqrt{5}}{2})J_{\perp}t}{\rm+}\frac{65{\rm-}3\sqrt{5}}{1800}\cos{(\frac{3{\rm-}\sqrt{5}}{2})J_{\perp}t}\\&+\frac{1}{450}\cos{2\sqrt{5}J_{\perp}t}-\frac{2}{45}\cos{(\sqrt{5}J_{\perp}t)}=\tau^{0zzz0}_{3,i}(t;5)
         \end{split}
 \end{equation}
 \begin{equation}
 \label{eq:RC}
     \begin{split} 
         \tau^{z0zz}_{3,i}(t;5)=&{\rm-}\frac{1}{90}{\rm-}\frac{3{\rm+}\sqrt{5}}{200}\cos{(\frac{5{\rm+}\sqrt{5}}{2})J_{\perp}t}{\rm-}\frac{3{\rm-}\sqrt{5}}{200}\cos{(\frac{5{\rm-}\sqrt{5}}{2})J_{\perp}t}\\&{\rm+}\frac{5{\rm-}3\sqrt{5}}{900}\cos{3(\frac{1{\rm-}\sqrt{5}}{2})J_{\perp}t}{\rm+}\frac{5{\rm+}3\sqrt{5}}{900}\cos{3(\frac{1{\rm+}\sqrt{5}}{2})J_{\perp}t}\\&{\rm+}\frac{65{\rm-}3\sqrt{5}}{1800}\cos{(\frac{3{\rm+}\sqrt{5}}{2})J_{\perp}t}{\rm+}\frac{65{\rm+}3\sqrt{5}}{1800}\cos{(\frac{3{\rm-}\sqrt{5}}{2})J_{\perp}t}\\&+\frac{1}{450}\cos{2\sqrt{5}J_{\perp}t}+\frac{2}{45}\cos{(\sqrt{5}J_{\perp}t)}=\tau^{zz0z}_{3,i}(t;5)
     \end{split}
 \end{equation}
\subsection*{Disconnected XX-network}
 \label{sec:disconnected}
 The last dynamical map components we determine are those associated to the disconnected XX network presented in Section 4. The eigen-parameterization used for $H^{1,2}_{\rm XX}$ is 
\begin{equation}
    \begin{split}
&h_{12}=\frac{h_{1}+h_{2}}{2} \quad \omega_{12}={\rm sgn}(\Delta_{12})\sqrt{\Delta_{12}^{2}+J_{12}^{2}}\\
&\Delta_{12}=h_{1}-h_{2} \quad \tan\phi_{12}=\frac{J_{12}}{\Delta_{12}}
        \end{split}
\end{equation}
Using these parameters, the Pauli components of the unitary channel may be expressed as
\begin{equation}
    \begin{split}
&\mathcal{U}^{x0}_{zy}=\mathcal{U}^{zx}_{y0}=\mathcal{U}^{0x}_{yz}=\mathcal{U}^{xz}_{0y}=-\mathcal{U}^{y0}_{zx}=-\mathcal{U}^{zy}_{x0}=-\mathcal{U}^{0y}_{xz}=-\mathcal{U}^{yz}_{0x}=\sin{\phi_{12}}\sin{\omega_{12} t}\cos{2 h_{12} t} \\  &\mathcal{U}^{xx}_{0z}=\mathcal{U}^{yy}_{z0}=\mathcal{U}^{z0}_{xx}=\mathcal{U}^{z0}_{yy}=-\mathcal{U}^{xx}_{0z}=-\mathcal{U}^{yy}_{0z}=-\mathcal{U}^{0z}_{xx}=-\mathcal{U}^{0z}_{yy}=\cos{\phi_{12}}\sin{\phi_{12}}\sin^{2}{\omega_{12} t} \\&\mathcal{U}^{xy}_{z0}=\mathcal{U}^{yx}_{z0}=\mathcal{U}^{z0}_{yx}=\mathcal{U}^{0z}_{xy}=-\mathcal{U}^{xy}_{0z}=-\mathcal{U}^{yx}_{z0}=-\mathcal{U}^{z0}_{xy}=-\mathcal{U}^{0z}_{yx}=\frac{1}{2}\sin{\phi_{12}}\sin{2\omega_{12} t}\\&\mathcal{U}^{0x}_{xz}=\mathcal{U}^{0y}_{yz}=\mathcal{U}^{xz}_{0x}=\mathcal{U}^{yz}_{0y}=\mathcal{U}^{x0}_{zx}=\mathcal{U}^{y0}_{zy}=\mathcal{U}^{zx}_{x0}=\mathcal{U}^{zy}_{y0}=\sin{\phi_{12}}\sin{\omega_{12} t}\sin{2 h_{12} t} \\&\mathcal{U}^{y0}_{x0}=-\mathcal{U}^{x0}_{y0}=\mathcal{U}^{yz}_{xz}=-\mathcal{U}^{xz}_{yz}=\cos\omega_{12} t\sin2h_{12}t+\cos{\phi_{12}}\sin\omega_{12} t\cos{2h_{12}t}  \\&\mathcal{U}^{0y}_{0x}=-\mathcal{U}^{0x}_{0y}=\mathcal{U}^{zy}_{zx}=-\mathcal{U}^{zx}_{zy}=\cos\omega_{12} t\sin2h_{12}t-\cos{\phi_{12}}\sin\omega_{12} t\cos{2h_{12}t} \\ &\mathcal{U}^{x0}_{x0}=\mathcal{U}^{y0}_{y0}=\mathcal{U}^{xz}_{xz}=\mathcal{U}^{yz}_{yz}=\cos\omega_{12} t\cos2h_{12}t-\cos{\phi_{12}}\sin\omega_{12} t\sin{2h_{12}t} 
\end{split}
\end{equation}
with the remaining components of the unitary channel given by
\begin{equation}
\begin{split}
&\mathcal{U}^{0x}_{0x}=\mathcal{U}^{0y}_{0y}=\mathcal{U}^{zx}_{zx}=\mathcal{U}^{zy}_{zy}=\cos\omega_{12} t\cos2h_{12}t{\rm+}\cos{\phi_{12}}\sin\omega_{12} t\sin{2h_{12}t} \\&\mathcal{U}^{yx}_{xx}=\mathcal{U}^{yy}_{xy}=-\mathcal{U}^{xx}_{yx}=-\mathcal{U}^{xy}_{yy}=\frac{1}{2}\big(\sin{4h_{12}t}{\rm+}\cos\phi_{12}\sin{2\omega_{12} t}\big)\\&\mathcal{U}^{xy}_{xx}=\mathcal{U}^{yy}_{yx}=-\mathcal{U}^{xx}_{xy}=-\mathcal{U}^{yx}_{yy}=\frac{1}{2}\big(\sin{4h_{12}t}-\cos\phi_{12}\sin{2\omega_{12} t}\big) \\  &\mathcal{U}^{yy}_{xx}=\mathcal{U}^{xx}_{yy}=\frac{1}{2}\big(-\cos{4h_{12}t}+\cos^{2}\phi_{12} \cos2\omega_{12} t+\sin^{2}\phi_{12} \big) \\
&\mathcal{U}^{xx}_{xx}=\mathcal{U}^{yy}_{yy}=\frac{1}{2}\big(\cos{4h_{12}t}+\cos^{2}\phi_{12} \cos2\omega_{12} t+\sin^{2}\phi_{12} \big) \\ &\mathcal{U}^{xy}_{xy}=\mathcal{U}^{yx}_{yx}=\frac{1}{2}\big(\cos{4h_{12}t}+\cos{2\omega_{12} t}\big) \\&\mathcal{U}^{xy}_{yx}=\mathcal{U}^{yx}_{xy}=\frac{1}{2}\big(\cos{4h_{12}t}-\cos{2\omega_{12} t}\big)  \\ &\mathcal{U}^{z0}_{z0}=\mathcal{U}^{0z}_{0z}=1-\sin^{2}{\phi_{12}}\sin^{2}{\omega_{12} t} \\
&\mathcal{U}^{z0}_{0z}=\mathcal{U}^{0z}_{z0}=\sin^{2}{\phi_{12}}\sin^{2}{\omega_{12} t}
 \end{split}
\end{equation}
Constructing the 1-qubit dynamical map, a non-unital dynamical map is obtained as given in Section 2. But enforcing a noise-distribution over $\phi_{12}$ that is an even function, the noise-averaged dynamical map takes the form Eq.(\ref{eq:phasecovariant}).

\bibliographystyle{unsrt}
\bibliography{MapEnsembles} 
\end{document}